%% file: main.tex
\documentclass[11pt]{article}
\usepackage[utf8]{inputenc}
\usepackage{graphicx}
\usepackage{booktabs}
\usepackage{geometry}
\usepackage{amsmath}
\usepackage{adjustbox}
\usepackage{natbib}
\usepackage{longtable}

\graphicspath{{./}}

\title{Founder Backgrounds and Startup Funding: Evidence from Y Combinator}
\author{Rommin Adl\\Grinnell College\\adlrommi@grinnell.edu}
\date{December 2025}

\begin{document}

\begin{titlepage}
\maketitle

\textbf{Abstract:} While founder backgrounds account for less than 4\% of funding variation among Y Combinator startups, this suggests that other factors, such as industry trends and product innovation, may play a more significant role in funding outcomes. Using data on 4,323 YC companies from 2005-2024 merged with S\&P Global funding data, I estimate OLS regressions with batch year fixed effects on a regression sample of 2,113 companies. The coefficient on prior FAANG work experience is -0.251, indicating approximately 22\% less funding. However, this result is not robust, as it changes direction in further analyses, suggesting that FAANG experience may not be a reliable predictor of funding. The most robust finding is that startups within Y Combinator that consist of larger founding teams tend to raise more funding, with each additional co-founder associated with approximately 21\% more capital raised. While observable credentials such as prior FAANG work experience and top-tier education explain minimal variation in funding, the size of the founding team emerges as a more consistent predictor, highlighting the importance of team dynamics in securing capital. Unobserved factors like industry and product quality likely dominate funding decisions within this elite accelerator cohort.

\vspace{0.3in}

\textbf{Keywords:} startup financing, founder human capital, startup accelerators, venture capital, Y Combinator, team composition, founder characteristics, Series A funding

\end{titlepage}

% ============================================
% INTRODUCTION AND LITERATURE REVIEW
% ============================================
\input{introduction_literature_review.tex}

% ============================================
% DATA SECTION
% ============================================
\input{data_section_RESTRUCTURED_FINAL.tex}

% ============================================
% MODEL SECTION
% ============================================
\input{model_section.tex}

% ============================================
% RESULTS SECTION
% ============================================
\input{results_section.tex}

% ============================================
% CONCLUSION
% ============================================
\section{Conclusion}

This study examined whether founder backgrounds predict funding outcomes within an elite cohort of Y Combinator startups, where all companies receive similar accelerator resources and network access. The analysis reveals that founder characteristics explain remarkably little of the variation in funding outcomes, with observable founder backgrounds accounting for less than 4\% of funding differences. The most robust finding is that larger founding teams raise significantly more funding, with each additional founder associated with approximately 21\% more capital. In contrast, the association between FAANG experience and funding outcomes is fragile and reverses sign depending on specification choices, suggesting that temporal trends and unobserved factors dominate any potential founder background effects.

This work contributes to the entrepreneurship literature by being the first large-scale empirical analysis of within-accelerator variation in founder backgrounds and funding outcomes. While previous research has established that accelerator participation improves startup performance relative to non-accelerated firms, and that founder human capital matters for venture success more broadly, few studies have examined whether founder characteristics continue to drive outcomes when accelerator experience is held constant. The findings here suggest that within an elite accelerator cohort, founder backgrounds may be less predictive than previously thought. This aligns with recent work by \citet{ren2025}, who found that education effects on funding were weaker for Y Combinator startups than for non-accelerated companies, suggesting that YC's reputation may serve as a strong quality signal that reduces the marginal value of individual founder credentials.

Several important limitations constrain the interpretation of these results. The observational nature of the data precludes causal claims, and the extremely low R-squared values indicate that unobserved factors, particularly industry, product quality, and market timing, likely dominate funding decisions. The fragility of the FAANG finding, which flips from positive in descriptive statistics to negative after controlling for batch year, underscores the challenges of identifying founder background effects when temporal trends and selection processes are at play. Additionally, the matching process between YC and S\&P Global data resulted in only 62.2\% of companies being successfully matched, potentially introducing selection bias if unmatched companies systematically differ in their funding patterns or founder characteristics.

For practitioners, these findings suggest that founder backgrounds alone provide limited predictive power for funding outcomes within elite accelerator cohorts. Accelerators and investors may find that product-market fit, industry dynamics, and market timing matter more than individual founder credentials when evaluating companies that have already passed rigorous accelerator selection. For entrepreneurs, the results indicate that while team composition matters, individual pedigree, whether it's through FAANG experience or top-tier education, might actually be less determinative of funding success than often assumed. However, these implications should be interpreted with caution given the descriptive nature of the analysis and the substantial unexplained variation in funding outcomes.

Future research should address the data limitations identified here. Most critically, future entrepreneurship studies need better measures of industry, product quality, and market timing to control for these unobserved factors that likely drive most funding variation. Research leveraging exogenous variation in prior FAANG experience, such as policy changes affecting tech employment or shifts in accelerator selection algorithms, could help establish causality. Additionally, instrumental variables approaches or natural experiments can help identify the mechanisms through which founder backgrounds affect startups' funding outcomes. Also, panel data examining the timing of funding milestones, rather than just binary indicators, would provide additional insights into how founder characteristics relate to the speed and trajectory of fundraising success.

This analysis demonstrates the importance of controlling for temporal trends and selection processes when examining founder background effects, and highlights the substantial data limitations that constrain understanding of what drives startup funding outcomes. While founder backgrounds may matter less within elite accelerator cohorts than in the broader startup ecosystem, the complexity of funding decisions and the dominance of unobserved factors suggest that simple founder credential signals are insufficient for predicting venture success. The relationship between founder characteristics and funding outcomes appears to be highly context-dependent, shaped by accelerator reputation, market conditions, and factors that remain unmeasured in the available data.

% ============================================
% REFERENCES
% ============================================
\clearpage
\bibliographystyle{apalike}
\bibliography{references}

\end{document}

%% file: introduction_literature_review.tex
\section{Introduction}

In recent years, high-profile startups such as Airbnb and Dropbox reached multi-billion-dollar valuations after graduating from Y Combinator, most commonly known as YC, the world's leading startup accelerator, located in San Francisco, California. Yet for every Airbnb, there are many more YC graduates who cannot raise substantial funding or scale their businesses. This gap truly raises the following question: what causes the disparity in success between the most successful YC startups and the struggling ones? One major theory is that the differences in founders' backgrounds, e.g., education credentials, pre-startup experience at premier tech companies, and the composition of their founding teams, actually dictate the future of these startups after partaking in an accelerator. YC startup founders often have degrees from top universities like Stanford or prior work experience at Google and Meta as badges of credibility; likewise, investors and accelerators frequently underscore the significance of a strong founding team with a good story, as these investors not only bet on the companies but also on the people running them. This study asks whether such founder characteristics can predict a startup's fundraising success after finishing Y Combinator. Particularly, this research examines if indicators of founders' human capital (education level, industry experience, prior FAANG work experience) and team composition (e.g., number of co-founders) are associated with greater venture funding success following the YC accelerator program.

Understanding whether founder background matters for post-YC success is important in regards to both the theory and practice. In practical terms, accelerators and early-stage investors are able to make their selection of startups and advising/mentoring focus more efficient if they had a sense of which founder characteristics are linked to better performance. If, for example, prior experience at a large tech firm or a larger founding group increases the likelihood of acquiring a Series A round, then accelerators can train or recruit in that direction. For founders and entrepreneurs, having that awareness of desirable success drivers would guide professional development and talent acquisition. At the scholarly level, this paper addresses a gap in entrepreneurship research. Growing bodies of evidence consistently show that startup accelerator involvement has a positive impact on startup performance relative to non-accelerated firms across the board. Similarly, there is a significant body of evidence that links founders' human capital, such as education and prior work experience to venture performance in terms of capital/funding raised. Few, if any, studies have examined variation within an elite cohort of accelerator-backed companies as the drop-off of accelerators after YC in terms of prestige is steep. Even though YC startups receive the same resources and access to network through the program, they do not all level out in terms of outcome later on. Through examining variation across YC alumni, this work will explain whether founder-level drivers of outcome persist even within an elite, well-funded startup environment. Briefly, this paper will analyze if founders' education, experience, and team makeup can be fundraising outcome predictors post-Y Combinator years based on a new dataset of YC startups and their post-program funding success.

To address this question, I construct a dataset merging Y Combinator company records with S\&P Global funding data, covering 4,323 companies from 2005-2024. I estimate OLS regressions with batch year fixed effects, examining whether founder characteristics---such as prior FAANG work experience, team size, and educational background---predict total funding raised. In a regression of log funding on founder characteristics controlling for batch year, the coefficient on FAANG experience is -0.251 (p=0.057), suggesting approximately 22\% less funding. However, this result is fragile and reverses sign in my robustness checks. The only robust finding is that larger founding teams raise more funding, with each additional co-founder associated with roughly 21\% more capital. Overall, founder backgrounds explain less than 4\% of funding variation.

This paper contributes to the entrepreneurship literature by being the first large-scale empirical analysis of within-accelerator variation in founder backgrounds and funding outcomes. While previous research has established that accelerator participation improves startup performance relative to non-accelerated firms, and that founder human capital matters for venture success more broadly, few studies have examined whether founder characteristics continue to drive outcomes when accelerator experience is held constant. The results suggest that YC's reputation may serve as such a strong quality signal that individual founder credentials become less determinative of funding success within this elite cohort.

\section{Literature Review}

There is a rich and extensive literature across entrepreneurship and strategy on founder background influencing startup performance, and on the role of startup accelerators in supporting new ventures. Previous work consistently finds that founder human capital in the form of education, experience in the labor market, and related skills positively correlates with many indicators of startup performance. As an example, \citet{robinson1994} surveyed over 2,000 entrepreneurs and found that educational attainment and prior business background were associated with greater success in self-employment enterprises. Subsequent research has continued to validate the relationship between education and entrepreneurial success: \citet{hsu2007} illustrated how among technology startups, those with higher degrees or startup experience more often obtained venture capital financing, and on better terms. Similarly, \citet{colombo2010} studied high-tech start-ups in Italy and concluded that human capital of the founding team had a positive effect on firm growth; they specifically found an indirect effect in which highly educated, experienced entrepreneurs easily received venture funding, which further drove company expansion. This is in line with the belief that educated businesspeople and businesspeople with sound industry or management experience provide credibility and capabilities that investors find valuable, thus improving the fortunes of the business. Indeed, venture capitalists often use founder ``quality'' as a selector: \citet{baum2004} found that biotech startups led by groups with superior scientific qualifications and industrial experience were more likely to receive VC and ultimately performed better, suggesting that investors effectively ``pick winners'' by choosing founders with higher human capital. Typically, the literature establishes founder experience and education as important signals and resources that can help enhance the likelihood of a startup's success. While this literature is able to establish that founder human capital matters in terms of the general startup population, it does not examine whether these effects persist within elite, pre-selected, highly-vetted, cohorts. My paper addresses this gap by examining variation within Y Combinator, where all founders have already passed rigorous selection and are given similar support, allowing me to test whether credentials still differentiate outcomes when baseline quality is held constant. Unlike previous studies that compare founders across the entire startup ecosystem, I hold accelerator quality constant and examine whether said founder characteristics continue to matter when all companies have already passed YC's rigorous selection process, thus filling in a critical gap as to the understanding how founder credentials function within elite cohorts.

At the same time, a new wave of research examines specific startup accelerators and their impact on venture performance. Accelerators like Y Combinator offer seed capital, guidance, and even networking contacts with other YC Founders over the course of the program, as different studies conclude that this participation in a premier accelerator like YC, conveys advantages beyond the financial resources provided. Current large-scale evidence by \citet{assenova2024} shows that accelerator graduates, on average, perform more successfully than non-accelerated startups. In their article in Strategic Management Journal, these authors analyzed 8,580 startups across 408 accelerators worldwide and compared startups that participated in accelerator programs with similar startups that were also selected but did not participate. Assenova and Amit found that accelerated startups were 3.4\% more likely to obtain venture capital funding, raised \$1.8 million more capital on average, and achieved higher revenues and employment growth. These results affirm the prior research suggesting that top-performing accelerators are able to expedite key milestones across the board. For instance, \citet{hallen2014} reported that these graduates from leading accelerators (e.g., YC and Techstars) reached fund-raising and exit milestones much earlier than comparable startups outside of these accelerators. Likewise, \citet{winston2015} too found that top-tier accelerator startups reached their next round of capitalization earlier than otherwise comparable startups that had received funding from leading angel investor syndicates. These papers argue that accelerators provide a boost through enhanced companies' business models, validation of their pitches to investors, and exposure to networks of capital. It is evident that not all accelerators are made equal, as the positive effects are most pronounced for the top-rated programs such as YC and Techstars, while lower-rated accelerators like Venture24 Inc produce more mixed outcomes. Nonetheless, most studies discussed in my literature review find that accelerator participation improves startup performance, thus validating the idea that the advice and quality indicator provided by programs such as YC helps new firms perform well in the market, leading to higher success. These studies primarily compare accelerated versus non-accelerated companies, making it difficult to distinguish accelerator effects from founder selection effects. My paper differs by focusing on variation within a single accelerator cohort, holding accelerator resources constant and examining whether founder backgrounds explain the substantial variation in outcomes among YC graduates. By examining within-accelerator variation rather than between-accelerator comparisons, I can isolate the role of founder characteristics from accelerator program effects, addressing a limitation that previous research could not overcome due to their comparative design.

Though the evidence that follows shows founder characteristics and accelerator attendance as being both consequential, few researchers have examined how these interact with each other within a single accelerator's cohort. Most accelerator research controls for participants versus non-participants, so it is hard to know whether outcome differences are because of the accelerator program or the quality of founders who get into them. One recent work by \citet{ren2025} starts to fill in this gap by analyzing how startups send quality signals to investors. Focusing partly on accelerator-backed companies, \citet{ren2025} show that the education level of a founder is an important signal that lowers perceived investment risk. They found that companies with founders that possess higher levels of education attracted greater amounts of overall funding, which is an impact in line with investors interpreting education as a good indicator of founder quality. Consequently, \citet{ren2025} found that this education effect on investment was larger for non-Y Combinator startups. One possible interpretation is that YC's reputation alone is such a strong quality signal that educational credentials within YC graduates are a less discriminating attribute among companies. This reinforces the importance of careful study of within-YC variability: while YC might award a quality certification to all of its startups, not all YC alumni perform as effectively at fundraising. To date, there is no specific analysis in academic work of which variations among founders in YC startups (or for that matter, any top accelerator cohort) are associated with better outcomes. The present study will fill this literature gap by examining the relative contribution of founders' human capital and team members to post-accelerator fundraising, holding accelerator experience constant for all firms in the sample. In the process, it aims to contribute new evidence on whether, even in a best-of-the-best group of startups, ``who the founders are'' continues to drive venture success. I build directly on Ren et al. by examining additional founder characteristics beyond education—specifically FAANG experience and team composition—and by using comprehensive funding data rather than focusing solely on education signals. This allows me to test whether the attenuation of credential effects within YC extends to work experience credentials as well. While Ren et al. found that education effects were weaker for YC startups, they did not examine whether this attenuation extends to work experience credentials like FAANG experience, nor did they analyze team composition effects. My analysis of FAANG experience and founder count provides a more comprehensive test of whether credential signals matter within elite accelerator cohorts, extending their finding beyond education to other dimensions of founder human capital.

%% file: data_section_RESTRUCTURED_FINAL.tex
% Add these packages at the top of your main document:
% \usepackage{graphicx}
% \usepackage{booktabs}

\section{Data}

\subsection{Data Sources}

This research study combines two primary data sources to examine the relationship between startup founder backgrounds and their companies' funding outcomes. The first data source is the Y Combinator Company and Founder Dataset (YC Dataset), obtained from Y Combinator's public directory and company records acquired through Kaggle. Y Combinator is one of the world's most prominent startup accelerators, having funded over 4,000 companies since its founding in 2005. The YC dataset includes information on companies accepted into YC batches from Winter 2005 (the first ever YC batch) through Summer 2024, representing 20 years of startups who have gone through the YC program. This dataset contains detailed founder-level characteristics including educational background and prior employment history, collected from public sources including LinkedIn profiles, company websites, and YC directory listings.

The second data source is S\&P Global Market Intelligence's private company funding database, accessed via their commercial data platform. S\&P Global delivers a thorough suite of funding data including transaction amounts and funding round types for private \& publicly traded companies, covering funding events from 2005 through September 2025. This ensures a strong coverage of funding events for companies in the sample period. Note that while S\&P Global records the date range of their data coverage, they do not provide transaction-level dates for individual funding rounds, which limits the ability to calculate timing-based metrics.

The data is cross-sectional at the company level, as each observation represents a single company. The funding data is at the transaction-level as there are multiple observations per company per multiple funding rounds, but are then aggregated to the company level for analysis. This allows me to create a single observation per company with total funding raised across all rounds. Additionally, research assistance was provided by Cursor \citep{cursor2025}.

\subsection{Sample Construction and Advantages}

This merged dataset provides a cross-sectional view of YC companies with available funding information. These companies serve as the unit of observation, with each characterized by founder background and aggregated funding outcomes. The company names were normalized using a standardized algorithm that removes common corporate suffixes, punctuation, and whitespace to allow for matching across datasets with potentially different naming conventions. 

The final merged sample includes 6,954 companies, of which 4,323 companies (62.2\%) have positive funding amounts recorded in the S\&P Global dataset. The full sample of 4,759 YC companies includes all companies regardless of funding data availability, while 4,599 companies (96.6\%) have at least some funding information recorded in the S\&P Global dataset. The regression analysis uses a further restricted sample of 2,113 companies that have non-missing values for all key variables (log total funding, FAANG experience indicator, founder count, and batch year). This sample restriction occurs because (1) companies with zero or missing funding amounts are excluded when using log-transformed funding as the dependent variable, and (2) companies with missing founder background data are excluded from the regression sample. The descriptive statistics tables report on different samples depending on variable availability, which explains why counts vary across tables.

This new merged dataset is highly relevant to my research question for several reasons. First, YC companies represent a relatively homogeneous population of early-stage technology startups that have passed a rigorous selection process, hence reducing concerns about unobserved heterogeneity that might confound founder background effects. Second, this combination of detailed founder-level characteristics such as education and employment history with the comprehensive funding data allows examination of the relationship between founder backgrounds and funding outcomes with greater precision than studies using aggregate data. Third, YC's prominence in the startup ecosystem means that the funding outcomes for YC companies are particularly relevant in understanding how investors evaluate founder quality signals as a metric for investment/funding. Fourth, YC's standardized application and selection process provides a relatively uniform baseline, additionally YC companies receive similar initial support and network access such as lawyers, previous YC founders, and venture capital firms, thus reducing variation in non-founder factors.

\subsection{Stylized Facts}

\subsubsection{Founder Characteristics}

The YC data provides multiple major founder characteristics that are relevant and leveraged in this analysis. Specifically prior employment at major technology firms is captured through an indicator variable for experience at FAANG companies (Facebook/Meta, Amazon, Apple, Netflix, and Google). The number of founders per company is also recorded, as team composition may actually influence funding outcomes. The average company in the sample has 1.90 founders, with a standard deviation of 0.80, ranging from 1 to 6 founders per company. Approximately 16.7\% of companies have at least one founder with prior FAANG experience.

These founder backgrounds exhibit substantial heterogeneity across a multitude of areas. Approximately 16.7\% of companies have at least one founder with prior FAANG experience, indicating that while FAANG experience is relatively common among YC founders, the majority of companies are founded by individuals without this specific background. This variation provides an identifying variation for examining how FAANG experience relates to funding outcomes. The founding team composition also varies substantially, as the average YC startup has 1.90 founders (standard deviation: 0.80), ranging from a single-founder to a founding team of up to 6 co-founders. This variation in team size may independently influence funding outcomes, as startups with larger founding teams might signal a greater commitment to the company, or even complementary skills with a technical cofounder and an entrepreneurial one, though they may also face coordination challenges and delayed execution.

Founder's educational backgrounds are categorized into four main levels: PhD, graduate degree (including MBA and Master's degrees), undergraduate degree, and other or missing. However, educational attainment shows limited variation across the sample due to data availability constraints. The analysis includes 4,028 companies (58.0\%) with education information classified, while 2,926 companies (42.1\%) have completely missing education information. Among companies with education data, virtually all fall into the ``other'' category (4,028 companies, or 100\% of those with education data), which includes companies with missing education data or founders with non-traditional educational backgrounds. The S\&P Global dataset does not contain a degree type variable in the schools sheet, which prevents classification into PhD, graduate, or undergraduate categories. This distribution reflects both the selective nature of YC acceptance and substantial data limitations in educational reporting. The high proportion of missing education data (42.1\%) and the lack of the more specific granular education categories e.g dropped out, suggest that these educational credentials are often not reported or not even available in structured form, which significantly constrains my ability to examine the educational effects. What this limitation actually means is that my regression analysis will primarily focus on FAANG experience and founder count as the main founder background variables, with education included as more of a control variable.

However, when examining technical and business education fields (derived from field of study rather than degree level), the data show that 38.2\% of companies have at least one founder with technical education, while 17.5\% have at least one founder with business education. This variation in educational fields, combined with the variation in FAANG experience and team composition, provides multiple dimensions along which founder backgrounds differ, providing avenues to examine how different aspects of founder backgrounds relate to funding outcomes.

Beyond the basic characteristics, what varies substantially is YC founders prior work experience, as the mean number of prior companies per founder is 4.13, with a standard deviation: 1.81, suggesting that most founders having worked at 2-6 prior companies before founding their YC company. Approximately 49.4\% of companies have at least one founder who previously worked at a YC company, suggesting that YC's network effects are substantial. Approximately 21.0\% of companies have founders with experience at top technology companies beyond FAANG, while 9.9\% have founders with consulting or finance experience. This heterogeneity in these founders work experience suggests that YC founders often bring a diverse set of prior work experiences to their startups, which are able to impact investors evaluations of their startups.

Funding outcomes are measured using two primary variables from the S\&P Global dataset. Total funding raised (in USD) aggregates all funding rounds for each company, providing a measure of overall capital acquisition. Funding amounts were converted from thousands to dollars where necessary and capped at \$500 million per individual round to prevent extreme outliers from skewing the analysis. The log transformation of total funding is used in regression analysis to address the highly right-skewed distribution of funding amounts. I also added a binary indicator to capture whether these companies successfully raised a Series A funding round at any point in time. This serves as a milestone-based success metric, as a company either raises Series A funding or does not. Furthermore the S\&P Global dataset does not actually contain any transaction-level dates for individual funding rounds, so the timing-based metrics such as ``Getting a Series A within 36 months'' cannot be calculated. Instead, a binary indicator for whether companies raised any Series A funding is used, which still provides a meaningful measure of funding success.

These funding outcomes exhibit substantial heterogeneity across all YC companies, mirroring the highly skewed nature of venture capital outcomes, with a winner takes all approach, where the mean total funding is \$23.36 million, while the median is \$2.15 million, indicating that a small number of companies raise substantially larger amounts than typical firms. This maximum total funding in the sample is \$500.50 million (after capping), while the 75th percentile is \$10.1 million and the 90th percentile is \$45.8 million. This wide distribution suggests that funding outcomes are highly heterogeneous, with most companies raising modest amounts while a few raise substantial sums. The log transformation of total funding has a mean of 14.58 and a standard deviation of 2.29, indicating a more symmetric distribution suitable for regression analysis.

\begin{figure}[htbp]
\centering
\includegraphics[width=0.8\textwidth]{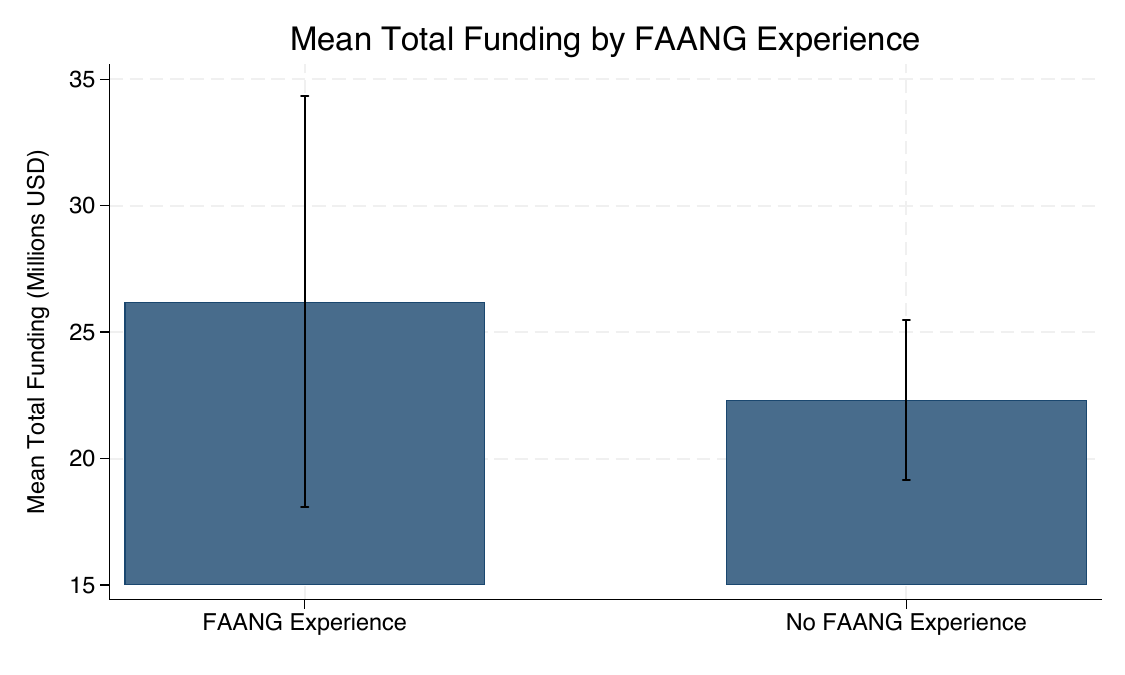}
\caption{Mean Total Funding by FAANG Experience}
\label{fig:funding_faang}
\textit{Notes:} This figure compares mean total funding between companies with and without FAANG-experienced founders. Error bars show 95\% confidence intervals.
\end{figure}

\textbf{FAANG Experience and Funding.} Companies with founders who have prior work experience in FAANG raise an average of \$26.21 million (N=375), while those without FAANG experience raise \$22.31 million (N=1,738), representing a difference of approximately 17.5\%. This pattern suggests that founders with prior FAANG work experience may be associated with decently higher funding outcomes. As Figure~\ref{fig:funding_faang} visualizes this comparison with 95\% confidence intervals, showing that while these FAANG experienced founders raise slightly more on average, their confidence intervals overlap substantially, indicating that the difference is not statistically significant in the raw data. This pattern motivates the regression analysis, which controls for other factors that may confound this relationship.

\begin{figure}[htbp]
\centering
\includegraphics[width=0.8\textwidth]{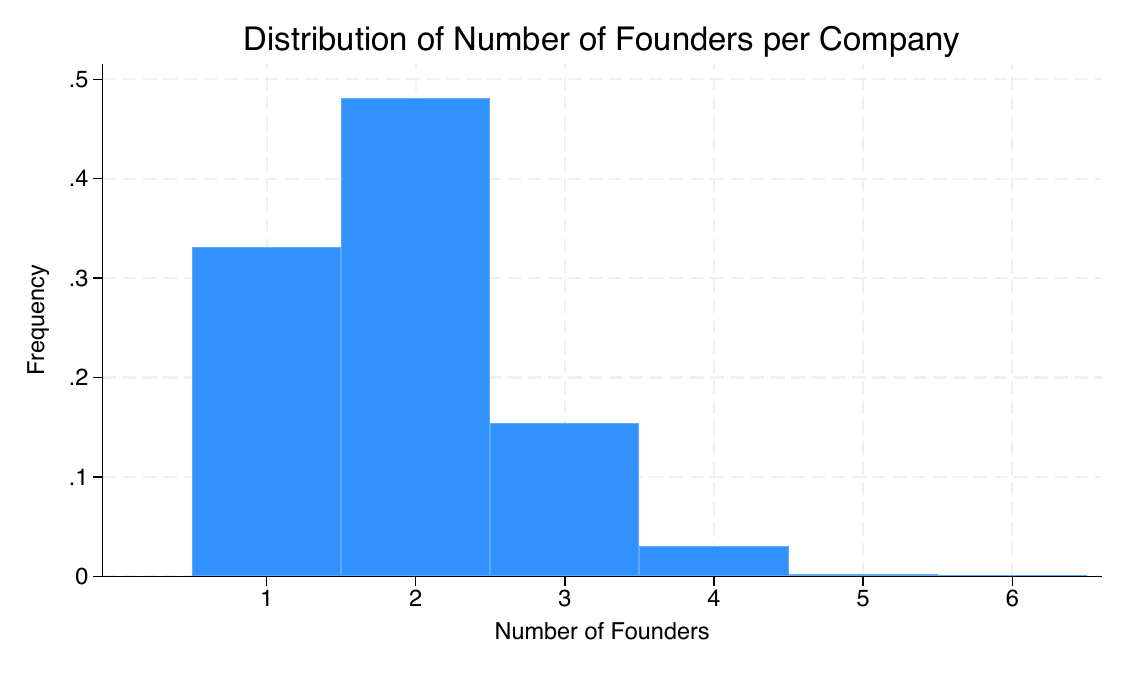}
\caption{Distribution of Number of Founders per Company}
\label{fig:founder_count}
\textit{Notes:} This figure shows the distribution of the number of founders per company across all YC companies in the sample.
\end{figure}

\textbf{Team Composition and Funding.} The average company has 1.90 founders (standard deviation: 0.80), with companies ranging from solo founders to most commonly duo co-founders and even to teams of up to 6 co-founders. Figure~\ref{fig:founder_count} shows the distribution of founder count, revealing that most companies have 1-2 founders. Team size may affect funding outcomes through several channels, as larger teams may signal greater commitment and complementary skills, but may also face coordination challenges. Founder count is controlled for in the regression analysis to isolate the effect of founder backgrounds from team size effects.

\begin{figure}[htbp]
\centering
\includegraphics[width=0.8\textwidth]{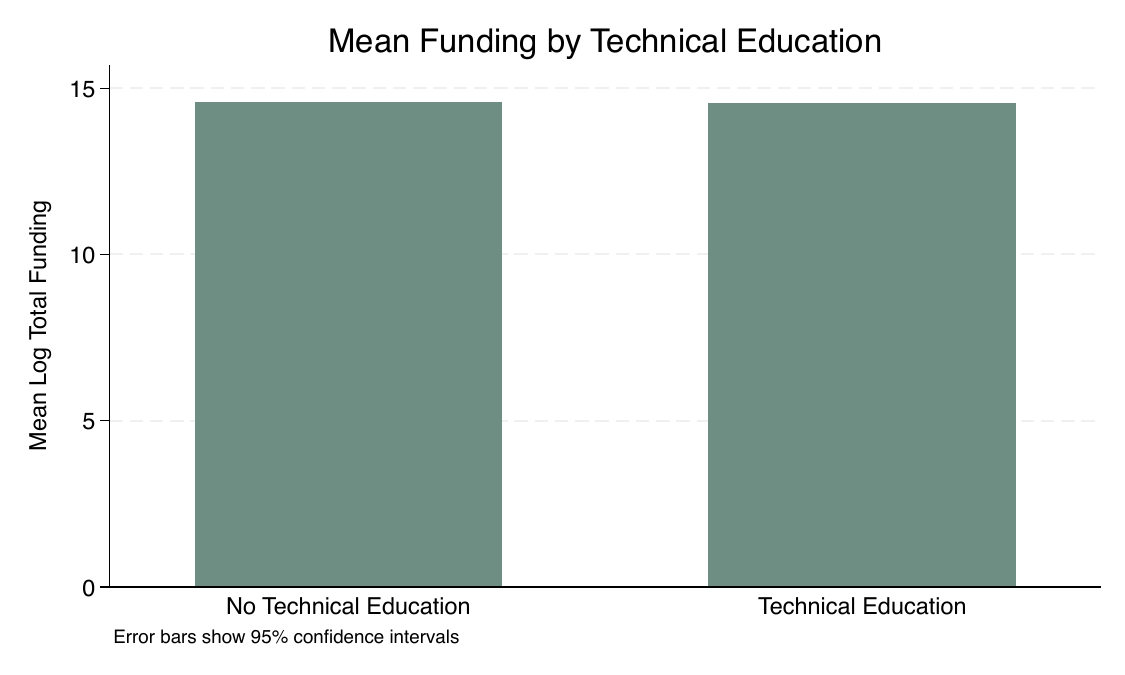}
\caption{Mean Funding by Technical Education}
\label{fig:funding_technical_ed}
\textit{Notes:} This figure compares mean log total funding between companies with and without founders who have technical education backgrounds. Error bars show 95\% confidence intervals.
\end{figure}

\begin{figure}[htbp]
\centering
\includegraphics[width=0.8\textwidth]{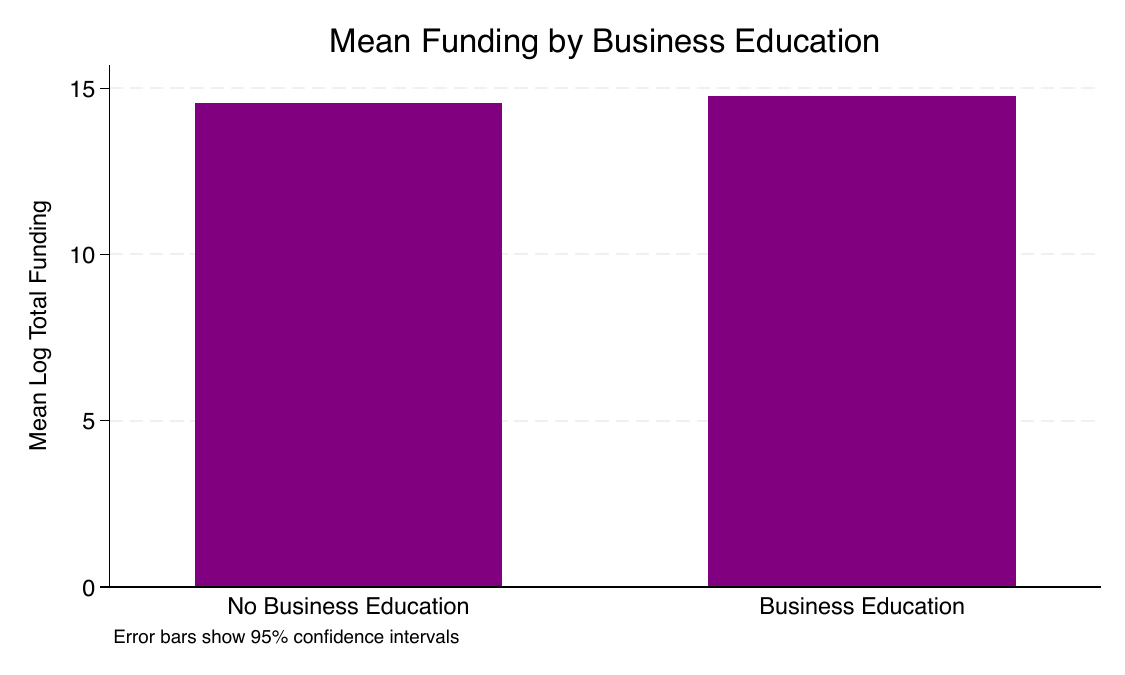}
\caption{Mean Funding by Business Education}
\label{fig:funding_business_ed}
\textit{Notes:} This figure compares mean log total funding between companies with and without founders who have business education backgrounds. Error bars show 95\% confidence intervals.
\end{figure}

\textbf{Education and Funding.} Founder education is classified into technical fields (computer science, software engineering, electrical engineering, mathematics, physics, statistics, data science, machine learning, artificial intelligence, cybersecurity, information systems) and business fields (business, MBA, finance, economics, accounting, marketing, management, entrepreneurship). Approximately 38.2\% of companies (1,651 out of 4,323) have at least one founder with technical education, while 17.5\% (756 out of 4,323) have at least one founder with business education. Figure~\ref{fig:funding_technical_ed} shows that companies with technical education raise slightly less funding on average (log funding: 14.57 versus 14.59), though the difference is very small and not statistically significant. Figure~\ref{fig:funding_business_ed} shows that companies with business education raise slightly more funding on average (log funding: 14.66 versus 14.57), though again the difference is modest. These patterns suggest that educational background may matter for funding outcomes, though the effects appear to be smaller than for work experience.

\begin{figure}[htbp]
\centering
\includegraphics[width=0.8\textwidth]{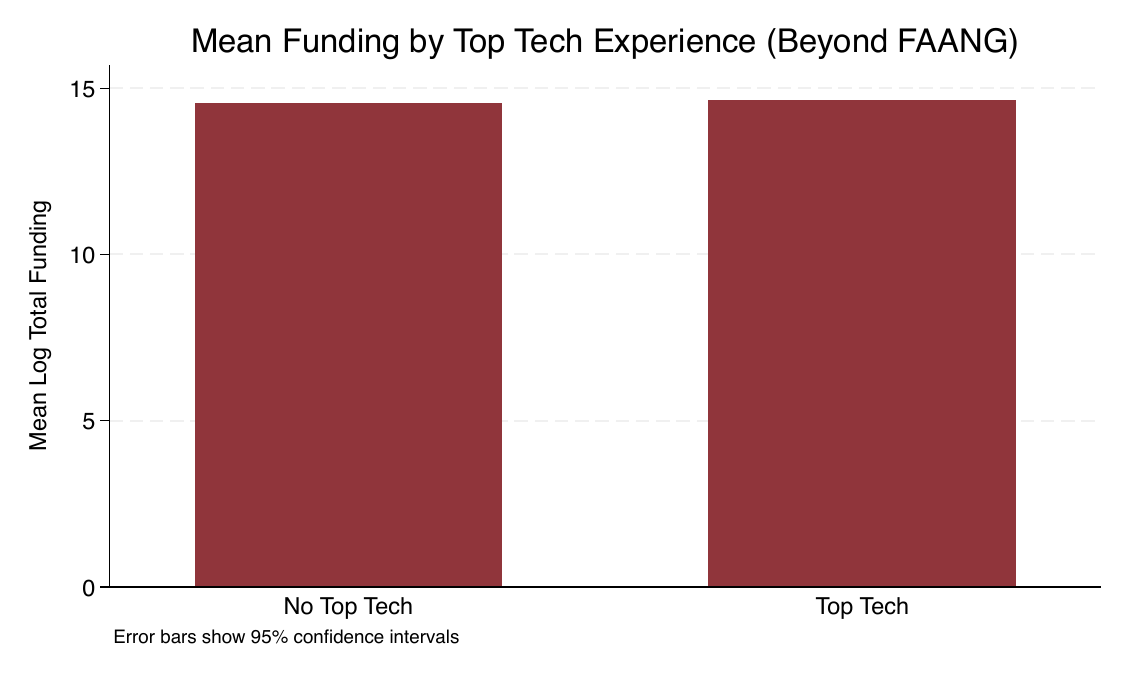}
\caption{Mean Funding by Top Tech Experience (Beyond FAANG)}
\label{fig:funding_top_tech}
\textit{Notes:} This figure compares mean log total funding between companies with and without founders who worked at top technology companies beyond FAANG. Error bars show 95\% confidence intervals.
\end{figure}

\begin{figure}[htbp]
\centering
\includegraphics[width=0.8\textwidth]{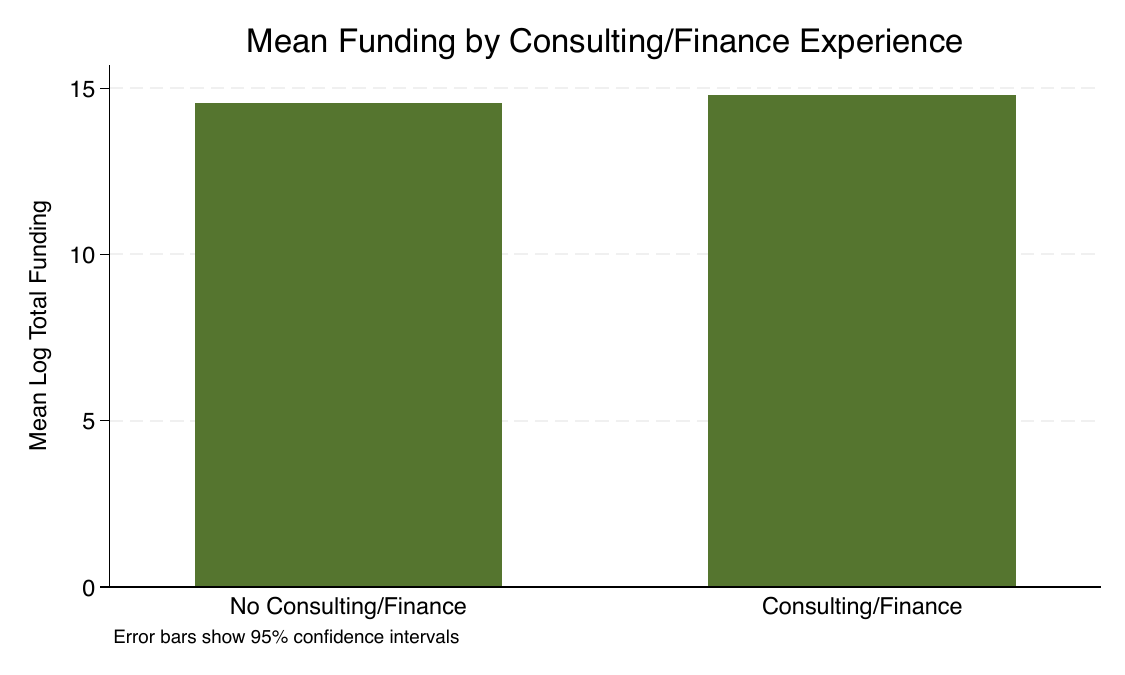}
\caption{Mean Funding by Consulting/Finance Experience}
\label{fig:funding_consulting}
\textit{Notes:} This figure compares mean log total funding between companies with and without founders who have consulting or finance experience. Error bars show 95\% confidence intervals.
\end{figure}

\textbf{Other Work Experience and Funding.} Beyond FAANG, approximately 21.0\% of companies (906 out of 4,323) have at least one founder with experience at top technology companies (Microsoft, Oracle, Salesforce, Adobe, Intel, IBM, Cisco, NVIDIA, and others, as well as prominent startups such as Stripe, Square, Uber, Airbnb). Figure~\ref{fig:funding_top_tech} shows that companies with top tech experience (beyond FAANG) raise slightly more funding on average (log funding: 14.64 versus 14.57), though the difference is modest and not statistically significant in the raw data. Approximately 9.9\% of companies (426 out of 4,323) have at least one founder with consulting or finance experience (McKinsey, Bain, BCG, Deloitte, PwC, Goldman Sachs, JPMorgan, Morgan Stanley, Blackstone, KKR, Sequoia, Kleiner Perkins, Andreessen Horowitz). Figure~\ref{fig:funding_consulting} shows that companies with consulting/finance experience raise substantially more funding on average (log funding: 14.80 versus 14.56), representing approximately 27\% more funding. This is one of the largest differences observed in the raw data, suggesting that consulting and finance experience may be particularly valuable for fundraising, possibly because these backgrounds provide strong analytical skills, network connections, or credibility with investors. However, this difference may also reflect selection into consulting/finance roles or other unobserved characteristics, which are addressed in the regression analysis.

Temporal variation in funding amounts is also substantial. The mean total funding varies substantially across YC batch years from 2005 through 2024, reflecting changes in overall venture capital market conditions over this two-decade period. Some periods show higher average funding than others, likely reflecting factors such as the post-2008 recovery, the 2020-2021 funding boom, and evolution in YC's selection criteria over the years. This temporal variation suggests that by controlling for batch year, I am able to avoid confounding founder background effects with changes in market conditions or YC selection criteria.

The heterogeneity in both founder characteristics and funding outcomes, combined with the substantial variation within groups (high standard deviations), suggests that founder backgrounds are actually not the sole determinant of funding success. Other factors such as product-market fit, market timing, industry, or unobserved company characteristics likely play important roles in this process. This heterogeneity motivates the empirical strategy to control for observable confounders and to acknowledge that unobserved factors may also influence funding outcomes. The variation in founder backgrounds (16.7\% with FAANG experience, 21.0\% with top tech experience, 9.9\% with consulting/finance experience) provides identifying variation for examining the relationship between founder backgrounds and funding outcomes, while the heterogeneity in funding outcomes (mean: \$23.36M, median: \$2.15M) suggests that there is meaningful variation to explain. The difference between the merged sample size (6,954) and the YC sample size (4,759) reflects companies that appear in the S\&P Global dataset but not in the YC companies list, which may represent companies that changed names or that the data is missing/incomplete.

\begin{figure}[htbp]
\centering
\includegraphics[width=0.8\textwidth]{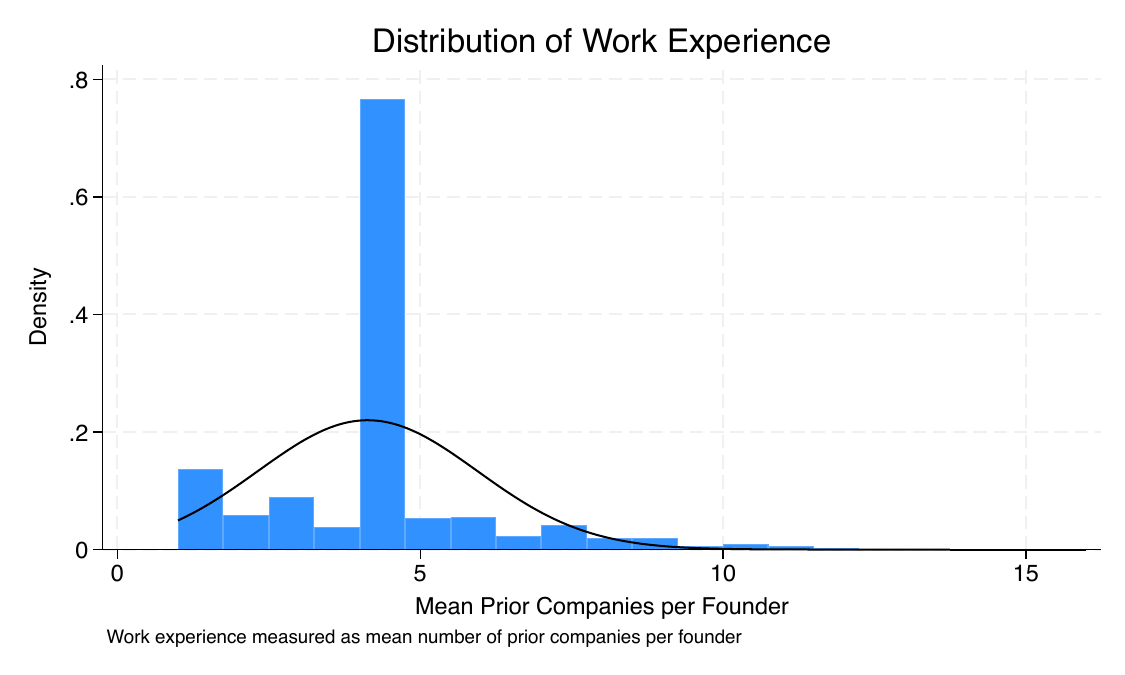}
\caption{Distribution of Work Experience (Mean Prior Companies per Founder)}
\label{fig:work_experience}
\textit{Notes:} This figure shows the distribution of mean number of prior companies per founder across all YC companies in the sample. Work experience is measured by counting the number of prior companies where each founder worked before founding their YC company. The mean is 4.13 prior companies per founder (standard deviation: 1.81), with most founders having worked at 2-6 prior companies.
\end{figure}

Beyond these basic characteristics, additional founder background variables are extracted from the YC dataset to provide a more comprehensive analysis of how founder backgrounds relate to funding outcomes. These variables capture dimensions of founder experience and education that may be relevant for investor evaluation but have not been systematically examined in previous research. These variables are constructed from the \texttt{prior\_companies}, \texttt{schools}, and \texttt{founders} sheets in the YC dataset, following a methodology similar to that used by Rebel Fund in their analysis of YC founder trends. Work experience is measured by counting the number of prior companies where each founder worked before founding their YC company. This variable captures the breadth of a founder's professional experience, which may signal adaptability, network connections, or exposure to different business models. The mean number of prior companies per founder is 4.13 (standard deviation: 1.81), indicating substantial variation in work experience across founders. Figure~\ref{fig:work_experience} shows the distribution of work experience, revealing that most founders have worked at 2-6 prior companies, with a small number having worked at 10 or more companies. This distribution suggests that while most founders have some prior work experience, there is meaningful variation that may be relevant for funding outcomes.

\begin{figure}[htbp]
\centering
\includegraphics[width=0.8\textwidth]{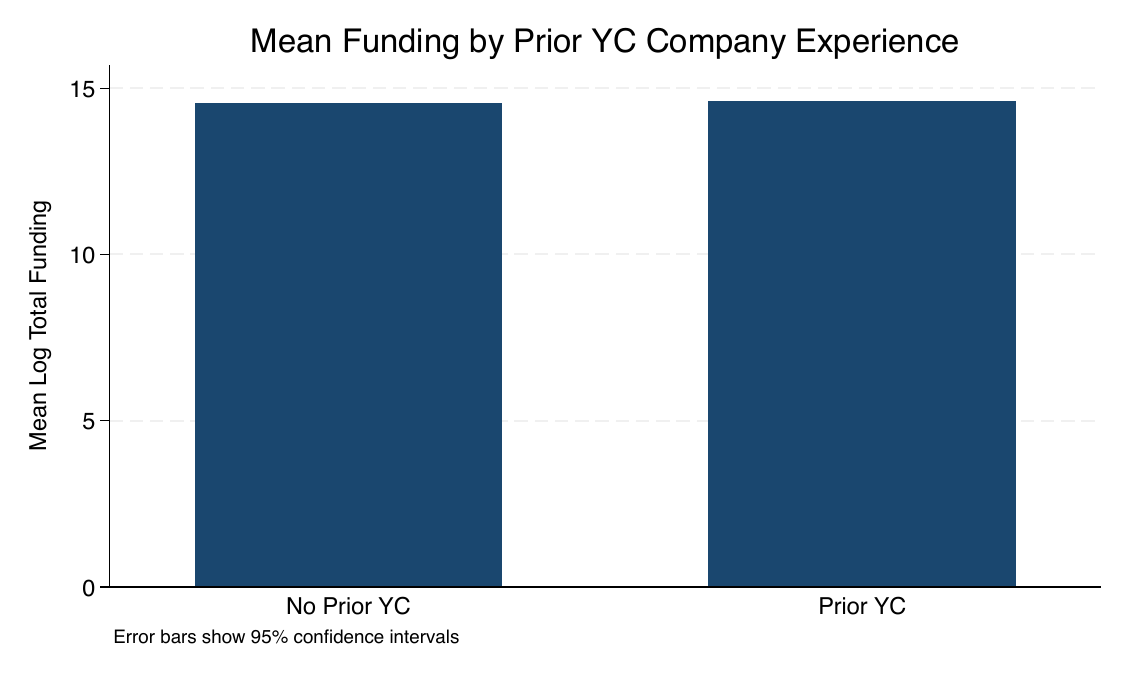}
\caption{Mean Funding by Prior YC Company Experience}
\label{fig:funding_prior_yc}
\textit{Notes:} This figure compares mean log total funding between companies with and without founders who worked at prior YC companies. Error bars show 95\% confidence intervals.
\end{figure}

Whether founders worked at prior YC companies is also identified, which may provide valuable network connections and insights into the YC process. Approximately 49.4\% of companies (2,135 out of 4,323) have at least one founder who previously worked at a YC company, suggesting that YC's network effects are substantial. Figure~\ref{fig:funding_prior_yc} compares mean funding between companies with and without founders who worked at prior YC companies. Companies with prior YC experience raise slightly more funding on average (log funding: 14.63 versus 14.54), though the difference is modest and the confidence intervals overlap substantially. This pattern suggests that prior YC experience may provide some advantage in fundraising, possibly through network connections or better understanding of investor expectations, though the effect appears to be small relative to other factors.

The analysis of prior employment is extended to include experience at other top technology companies beyond FAANG. This includes major technology firms such as Microsoft, Oracle, Salesforce, Adobe, Intel, IBM, Cisco, NVIDIA, and others, as well as prominent startups such as Stripe, Square, Uber, Airbnb, and others. Approximately 21.0\% of companies (906 out of 4,323) have at least one founder with experience at these top tech companies. Figure~\ref{fig:funding_top_tech} shows that companies with top tech experience (beyond FAANG) raise slightly more funding on average (log funding: 14.64 versus 14.57), though again the difference is modest and not statistically significant in the raw data. This pattern suggests that experience at top technology companies more broadly may provide similar signaling value to FAANG experience, though the effect appears to be smaller than for FAANG companies specifically.

Founders with experience in consulting (e.g., McKinsey, Bain, BCG, Deloitte, PwC) or finance (e.g., Goldman Sachs, JPMorgan, Morgan Stanley, Blackstone, KKR, Sequoia, Kleiner Perkins, Andreessen Horowitz) are also identified. Approximately 9.9\% of companies (426 out of 4,323) have at least one founder with consulting or finance experience. Figure~\ref{fig:funding_consulting} shows that companies with consulting/finance experience raise substantially more funding on average (log funding: 14.80 versus 14.56), representing approximately 27\% more funding (exp(0.24) - 1 $\approx$ 0.27). This is one of the largest differences observed in the raw data, suggesting that consulting and finance experience may be particularly valuable for fundraising, possibly because these backgrounds provide strong analytical skills, network connections, or credibility with investors. However, this difference may also reflect selection into consulting/finance roles or other unobserved characteristics, which are addressed in my regression analysis.

Founder education is classified into technical fields (computer science, software engineering, electrical engineering, mathematics, physics, statistics, data science, machine learning, artificial intelligence, cybersecurity, information systems) and business fields (business, MBA, finance, economics, accounting, marketing, management, entrepreneurship). Approximately 38.2\% of companies (1,651 out of 4,323) have at least one founder with technical education, while 17.5\% (756 out of 4,323) have at least one founder with business education. Figure~\ref{fig:funding_technical_ed} shows that companies with technical education raise slightly less funding on average (log funding: 14.57 versus 14.59), though the difference is very small and not statistically significant. Figure~\ref{fig:funding_business_ed} shows that companies with business education raise slightly more funding on average (log funding: 14.66 versus 14.57), though again the difference is modest. These patterns suggest that educational background may matter for funding outcomes, though the effects appear to be smaller than for work experience. The negative association with technical education is particularly interesting and may reflect that technical founders are more likely to bootstrap or pursue different funding strategies, or that investors value business skills more than technical skills in the YC context.

\begin{figure}[htbp]
\centering
\includegraphics[width=0.8\textwidth]{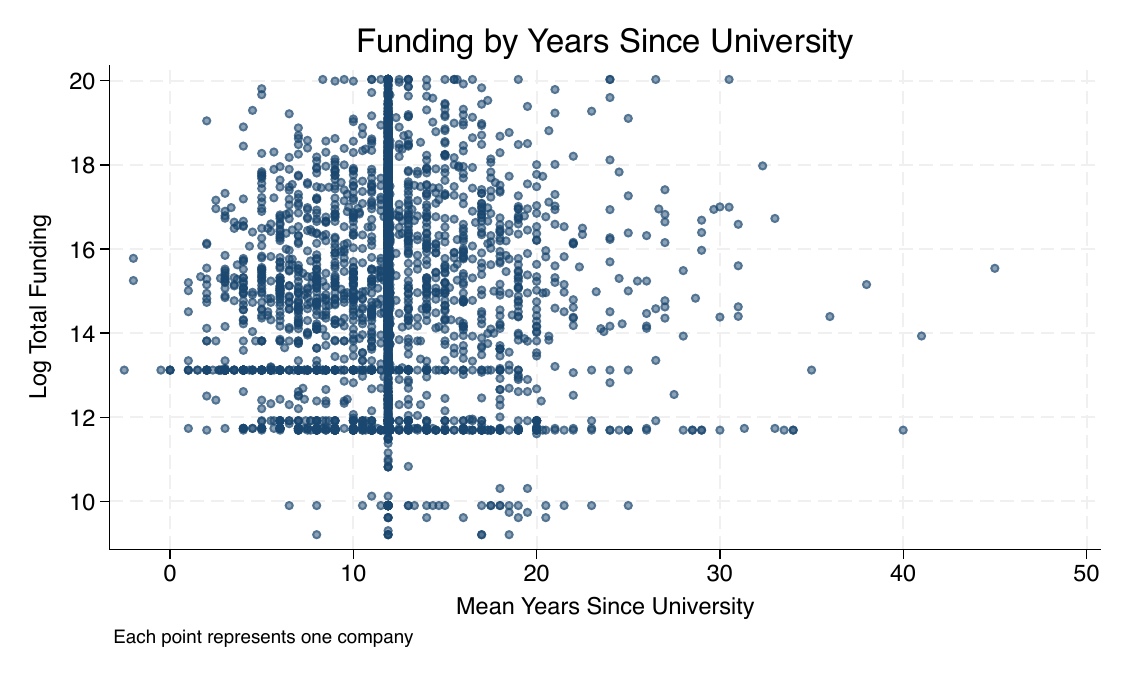}
\caption{Funding by Years Since University}
\label{fig:years_since_univ}
\textit{Notes:} This figure shows a scatter plot of log total funding against mean years since university graduation for each company. The line shows a fitted linear trend.
\end{figure}

Years since university graduation are calculated by identifying the maximum graduation year for each founder and subtracting it from 2025. This variable captures founder experience and maturity, which may be relevant for investor evaluation. The mean years since university is 11.9 (standard deviation: 4.1), indicating that most founders graduated approximately 10-15 years before founding their YC company. Figure~\ref{fig:years_since_univ} shows a scatter plot of funding by years since university, revealing a weak positive relationship between years since graduation and funding outcomes. This pattern suggests that more experienced founders (those who graduated longer ago) may raise slightly more funding, possibly because they have accumulated more skills, networks, or credibility over time. However, the relationship is not strong, and there is substantial variation around the trend line, suggesting that other factors are more important for funding outcomes.

\begin{figure}[htbp]
\centering
\includegraphics[width=0.8\textwidth]{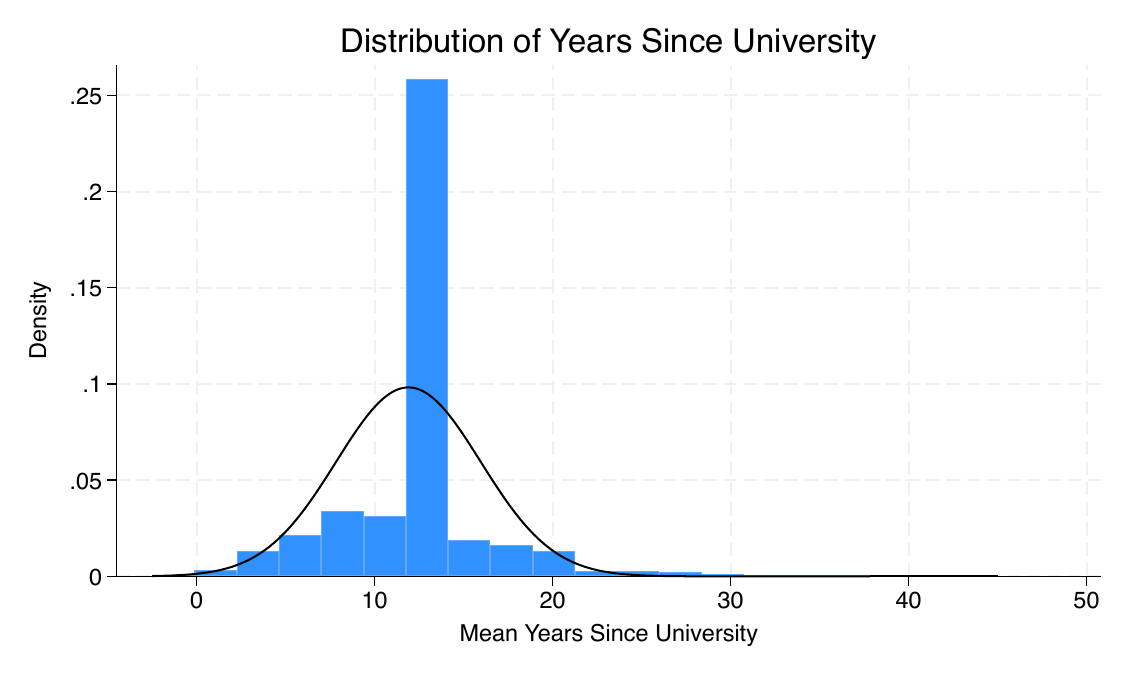}
\caption{Distribution of Years Since University}
\label{fig:years_distribution}
\textit{Notes:} This figure shows the distribution of mean years since university graduation across all YC companies in the sample.
\end{figure}

Figure~\ref{fig:years_distribution} shows the distribution of years since university, which is roughly normal with a mean of 11.9 years. This distribution suggests that most YC founders are in their early to mid-30s (assuming graduation at age 22), which is consistent with the typical age range for startup founders. The distribution shows substantial variation, with some founders having graduated very recently (less than 5 years ago) and others having graduated decades ago (more than 20 years ago). This variation provides identifying variation for examining how founder experience relates to funding outcomes.

\begin{figure}[htbp]
\centering
\includegraphics[width=0.8\textwidth]{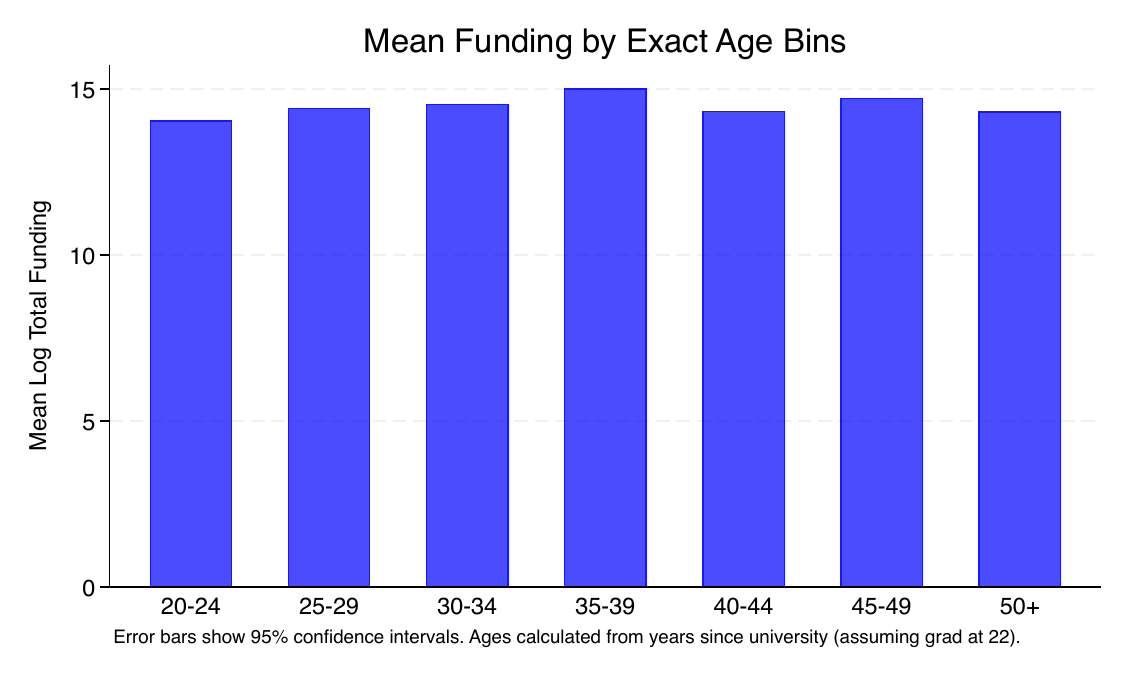}
\caption{Mean Funding by Exact Age Bins}
\label{fig:funding_exact_age}
\textit{Notes:} This figure shows mean log total funding by exact age bins (20-24, 25-29, 30-34, 35-39, 40-44, 45-49, 50+). Error bars show 95\% confidence intervals.
\end{figure}

Founder age is also constructed by adding 22 (the typical graduation age) to years since university. The mean founder age is 33.9 years (standard deviation: 6.0), with a range from 20 to 67 years. This age distribution is consistent with previous research on startup founders, which finds that successful founders are typically in their 30s. Age bins (20-24, 25-29, 30-34, 35-39, 40-44, 45-49, 50+) are created to examine how funding outcomes vary by founder age. Figure~\ref{fig:funding_exact_age} shows mean funding by exact age bins, revealing an interesting pattern: funding is lowest for the youngest founders (20-24), increases for the 25-29 age group (the peak age range identified by Rebel Fund), peaks for the 35-39 age group, and then declines for older founders. This pattern suggests that there may be an optimal age range for fundraising, with founders in their mid-30s raising the most funding on average. However, the confidence intervals are wide for some age groups (particularly the youngest and oldest), mirroring small sample sizes, so these patterns should be interpreted with caution.

\begin{figure}[htbp]
\centering
\includegraphics[width=0.8\textwidth]{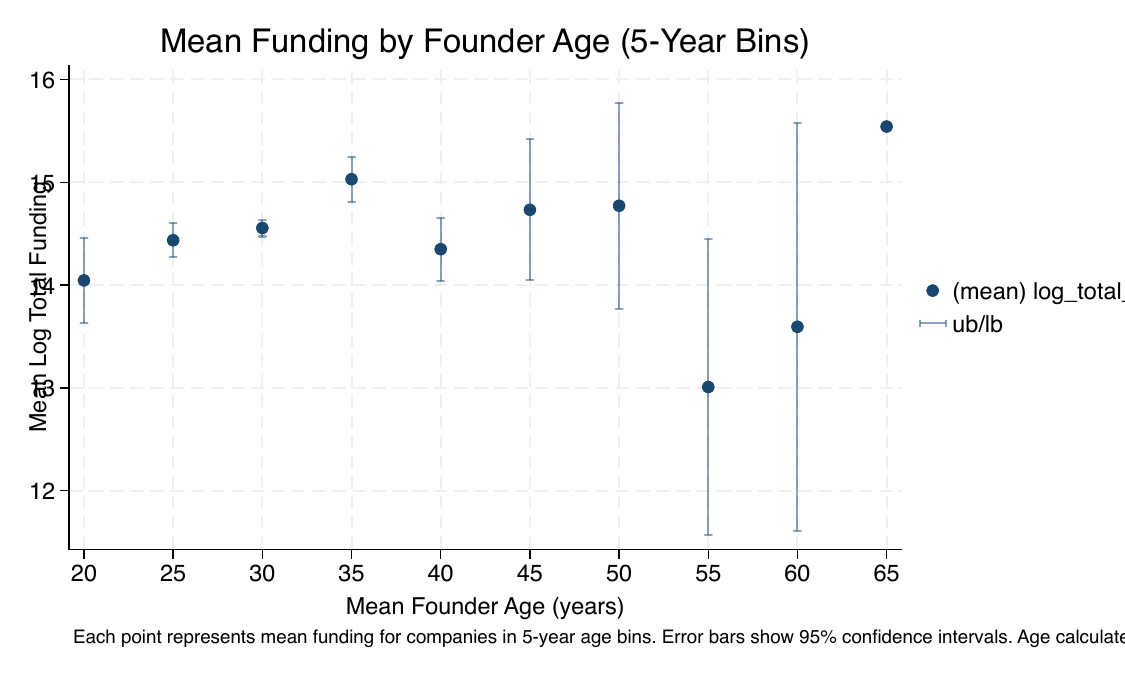}
\caption{Mean Funding by Founder Age (5-Year Bins)}
\label{fig:funding_exact_age_scatter}
\textit{Notes:} This figure shows a binned scatter plot of mean log total funding by mean founder age, grouped into 5-year bins. Error bars show 95\% confidence intervals.
\end{figure}

Figure~\ref{fig:funding_exact_age_scatter} shows a binned scatter plot of funding by founder age, grouping companies into 5-year age bins to better handle the fact that companies have multiple founders. Each point represents the mean funding for companies in that age bin, with error bars showing 95\% confidence intervals. This visualization confirms the pattern observed in Figure~\ref{fig:funding_exact_age}, showing that funding increases from the youngest age groups, peaks around age 35-39, and then declines for older founders. The binned scatter plot is more appropriate than a simple scatter plot because it accounts for the fact that each company aggregates multiple founders, and it shows the relationship more clearly by reducing noise from individual company variation.

\begin{figure}[htbp]
\centering
\includegraphics[width=0.8\textwidth]{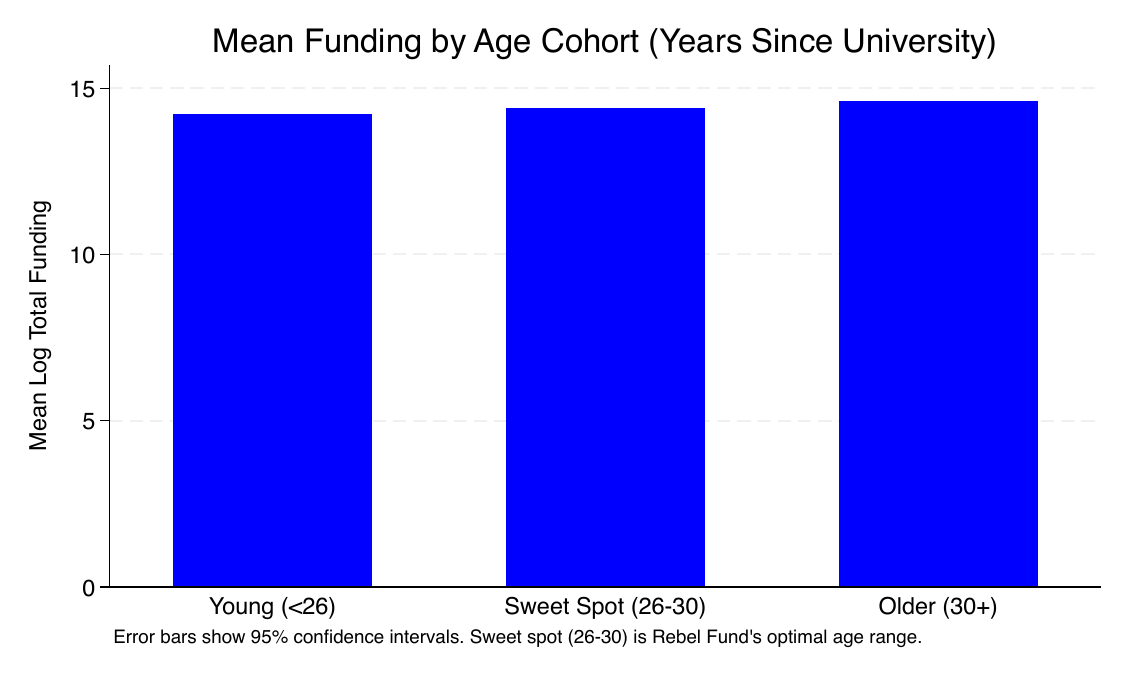}
\caption{Mean Funding by Age Cohort (Years Since University)}
\label{fig:funding_age_cohort}
\textit{Notes:} This figure shows mean log total funding by age cohort: young founders (under 26), sweet spot (26-30), and older founders (30+). Error bars show 95\% confidence intervals.
\end{figure}

Age cohorts are also created based on years since university to compare with Rebel Fund's analysis, which identified ages 26-30 as the optimal range for YC founders. Figure~\ref{fig:funding_age_cohort} shows mean funding by age cohort: young founders (under 26, assuming graduation at 22), the peak age range (26-30), and older founders (30+). The figure shows that funding is highest for older founders (30+), followed by the sweet spot (26-30), and lowest for young founders (under 26). This pattern differs from Rebel Fund's finding that the 26-30 age range is optimal, suggesting that in this sample, older founders raise more funding on average. However, this may reflect that older founders have more experience, networks, or other characteristics that are valued by investors, rather than age itself being the causal factor.

\input{table1_summary_stats.tex}

\begin{figure}[htbp]
\centering
\includegraphics[width=0.8\textwidth]{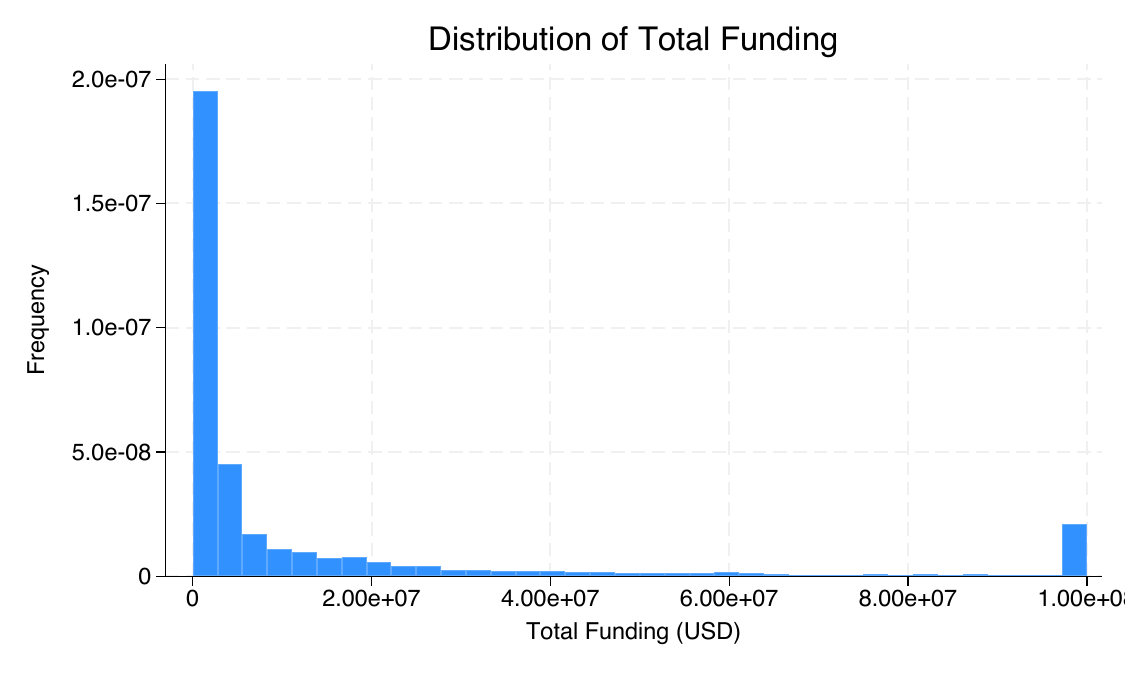}
\caption{Distribution of Total Funding}
\label{fig:funding_dist}
\textit{Notes:} This figure shows the distribution of total funding amounts across all YC companies in the sample. Values above \$100 million are capped for visualization purposes.
\end{figure}

Table~\ref{tab:summary_stats} presents summary statistics for the key variables. The average company has approximately 1.90 founders, with 16.7\% of companies having at least one founder with FAANG experience. Total funding amounts vary substantially, with a mean of \$23.36 million and a median of \$2.15 million. This reflects the right-skewed nature of venture funding, where a few companies raise significantly more than typical firms. The maximum total funding in the sample is \$500.50 million (after capping). The log transformation of total funding has a mean of 14.58 and a standard deviation of 2.29, indicating a more symmetric distribution suitable for regression analysis. The substantial variation in founder backgrounds (16.7\% with FAANG experience) provides identifying variation for the empirical strategy. The right-skewed distribution of funding amounts (Figure~\ref{fig:funding_dist}) motivates the log transformation used in the regression analysis, as it addresses the non-normal error distribution that would arise from using levels. The variation in FAANG experience across companies provides sufficient identifying variation to estimate the relationship between founder backgrounds and funding outcomes, though I note that education variation is limited due to data constraints.

Figure~\ref{fig:funding_dist} shows the distribution of total funding, displaying the characteristic right tail of venture capital outcomes where a few companies raise much larger amounts than typical firms. Values above \$100 million are capped for visualization purposes, but the underlying data shows substantial variation with the 75th percentile at \$10.1 million and the 90th percentile at \$45.8 million. This distributional pattern confirms that the log transformation is appropriate for regression analysis, as it will help address the heteroskedasticity and non-normality that would arise from the highly skewed distribution of funding amounts. The wide gap between the median (\$2.15 million) and mean (\$23.36 million) reflects the power-law distribution common in venture capital, where a small number of highly successful companies (``unicorns'') account for a disproportionate share of total funding. This distribution suggests that funding outcomes are highly heterogeneous, with most companies raising modest amounts while a few raise substantial sums, which motivates my use of log transformation to better capture the relationship between founder backgrounds and funding across this wide range of outcomes.

\begin{figure}[htbp]
\centering
\includegraphics[width=0.8\textwidth]{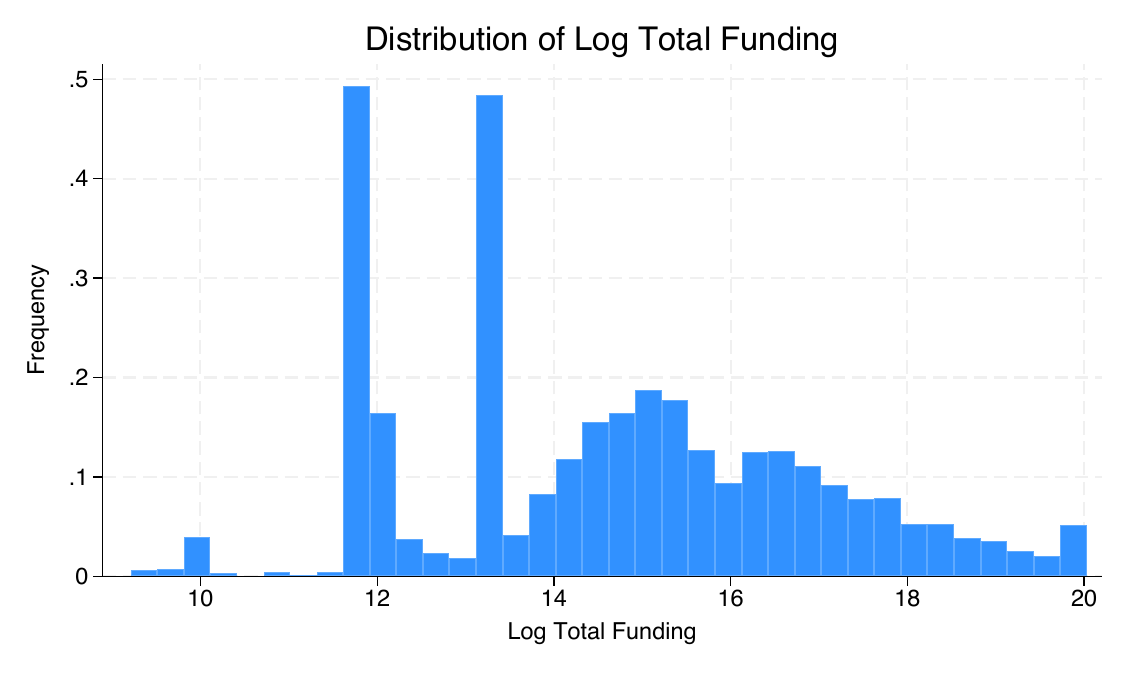}
\caption{Distribution of Log Total Funding}
\label{fig:log_funding}
\textit{Notes:} This figure shows the distribution of log total funding, which is more symmetric and better suited for regression analysis than the raw funding distribution.
\end{figure}

Figure~\ref{fig:log_funding} shows the distribution of log total funding, which is more symmetric and better suited for regression analysis, with a roughly normal distribution centered around 14.58. The log transformation successfully addresses the right-skewed nature of the funding distribution, making the error terms more likely to satisfy OLS assumptions of normality and homoskedasticity.

\input{table3_funding_by_faang.tex}

Table~\ref{tab:funding_faang} compares funding outcomes between companies with and without FAANG-experienced founders. Companies with FAANG founders raise an average of \$26.21 million (N=375), while those without FAANG experience raise \$22.31 million (N=1,738), representing a difference of approximately 17.5\%. This pattern suggests that FAANG experience may be associated with modestly higher funding outcomes, though the differences are relatively small and establishing causality requires careful identification. The standard deviations are substantial in both groups, indicating high within-group variation that may reflect differences in company age, industry, or other unobserved characteristics. This variation motivates the empirical strategy to control for these potential confounders in the regression analysis. The fact that the funding difference (\$3.9 million, or 17.5\% higher) suggests that FAANG experience may be more valuable for raising larger funding amounts, possibly mirroring investor preferences for founders with proven track records at scale.

Figure~\ref{fig:funding_faang} visualizes the mean funding by FAANG experience with 95\% confidence intervals. The figure shows that while FAANG-experienced founders raise slightly more on average (\$26.21 million versus \$22.31 million), the confidence intervals overlap substantially, indicating that the difference is not statistically significant in the raw data. This pattern motivates my regression analysis, which will control for other factors that may confound this relationship. The overlapping confidence intervals suggest that the raw difference may be driven by factors other than FAANG experience alone, such as company age, industry, or batch year effects. My regression analysis will help isolate the effect of FAANG experience by controlling for these potential confounders, allowing me to determine whether the observed difference reflects a causal relationship or is instead driven by selection into FAANG companies or other unobserved characteristics.

Figure~\ref{fig:founder_count} shows the distribution of founder count, revealing that most companies have 1-2 founders, with a small number having 3 or more. This distribution suggests that team composition varies across companies, which may influence funding outcomes independently of founder backgrounds. The mean of 1.90 founders per company indicates that most YC companies are founded by small teams, which is consistent with the early-stage nature of YC companies. Team size may affect funding outcomes through several channels: larger teams may signal greater commitment and complementary skills, but may also face coordination challenges or dilution of equity. Founder count is controlled for in the regression analysis to isolate the effect of founder backgrounds (FAANG experience) from team size effects, ensuring that estimates of FAANG effects are not confounded by differences in team composition.

\input{table4_funding_by_education.tex}

Table~\ref{tab:funding_education} presents funding outcomes stratified by founder education level. However, due to data limitations in the education variable (as discussed above), virtually all companies with education data fall into the ``other'' category, and 42.1\% of companies have missing education information. This lack of variation in education categories prevents meaningful comparisons across education levels.

\begin{figure}[htbp]
\centering
\includegraphics[width=0.8\textwidth]{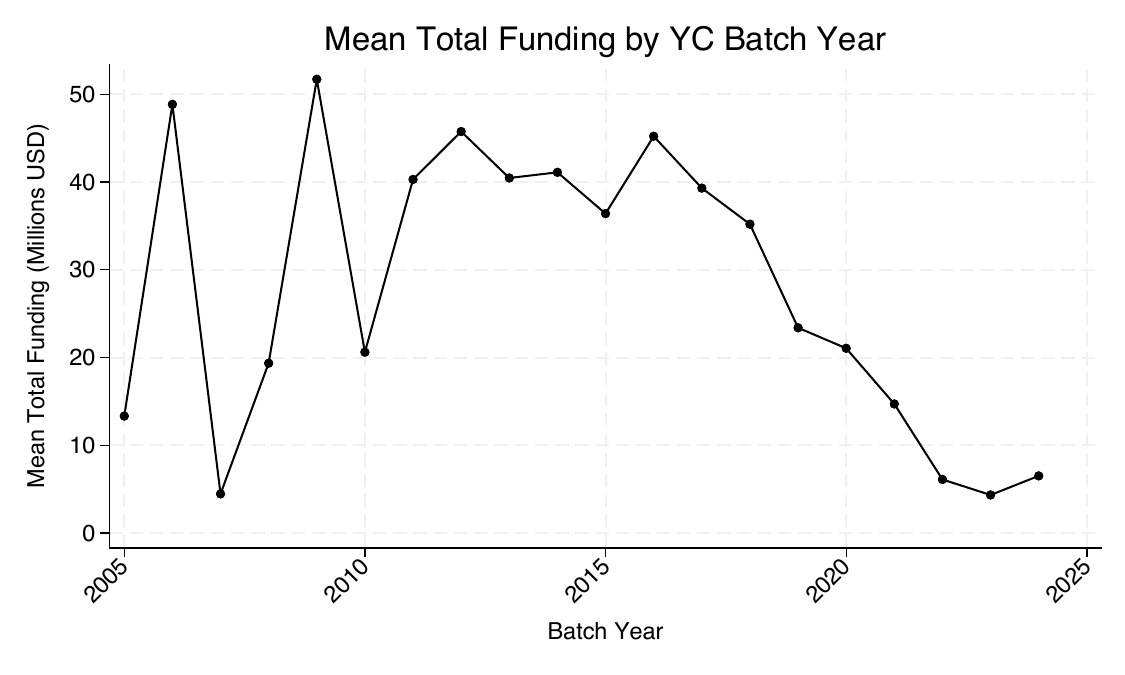}
\caption{Mean Total Funding by YC Batch Year}
\label{fig:funding_batch}
\textit{Notes:} This figure shows mean total funding by YC batch year from 2005 through 2024, revealing temporal patterns in funding amounts.
\end{figure}

Figure~\ref{fig:funding_batch} examines funding trends over time by plotting mean total funding by YC batch year from 2005 through 2024. This reveals temporal patterns in funding amounts, which may reflect changes in market conditions over the two-decade period and evolution in YC's selection criteria. The time series shows substantial variation across years, with some periods showing higher average funding than others. These temporal patterns likely reflect several factors: changes in overall venture capital market conditions (e.g., the post-2008 recovery, the 2020-2021 funding boom), evolution in YC's selection criteria and company quality over time, and changes in the types of companies that YC accepts. Batch year is controlled for in the regression analysis to account for these temporal trends and ensure that estimates of founder background effects are not confounded by changes in market conditions or YC selection criteria over time. This is particularly important because if YC's selection criteria have changed over time in ways that correlate with founder backgrounds (e.g., if YC increasingly values FAANG experience), failing to control for batch year could bias my estimates.

The descriptive patterns from the advanced founder characteristics suggest several key relationships that will inform my regression analysis. First, consulting and finance experience shows the strongest positive association with funding outcomes in the raw data, suggesting that these backgrounds may be particularly valuable for fundraising. Second, technical education shows a weak negative association with funding, which is surprising and may reflect different funding strategies or investor preferences. Third, founder age shows a non-linear relationship with funding, with funding peaking in the 35-39 age range. Fourth, prior YC experience and top tech experience show modest positive associations with funding, though the effects are smaller than for consulting/finance experience. These patterns motivate my regression analysis, which will control for other factors to isolate the causal effects of these founder characteristics on funding outcomes.

The descriptive patterns suggest several key relationships that motivate my empirical analysis. First, the variation in founder backgrounds (16.7\% with FAANG experience, 21.0\% with top tech experience, 9.9\% with consulting/finance experience) provides identifying variation for examining the relationship between founder backgrounds and funding outcomes. Second, the right-skewed distribution of funding amounts motivates the use of log transformation in the regression analysis, as it addresses the heteroskedasticity and non-normality that would arise from the highly skewed distribution. Third, the differences in funding by founder characteristics (e.g., consulting/finance experience showing the largest positive difference) suggest that founder backgrounds may influence funding success, though establishing causality requires careful identification. Fourth, the substantial variation in funding outcomes within groups (high standard deviations) suggests that founder backgrounds are not the sole determinant of funding success, and controlling for other factors is important. Fifth, the temporal variation in funding amounts across batch years (Figure~\ref{fig:funding_batch}) suggests that controlling for batch year is essential to avoid confounding founder background effects with changes in market conditions or YC selection criteria over time. These patterns inform my empirical strategy, which I describe in the Model section. I will control for batch year, founder count, and other company characteristics to isolate the effect of founder backgrounds on funding outcomes, while acknowledging that education effects cannot be examined due to data limitations.

In this paper there are several data limitations in particular the matching process between YC and S\&P Global relies on company name normalization, which may introduce measurement error if companies are listed under different names in the two sources or if the company has changed it's name overtime. Additionally companies that were not able to be matched are excluded from this analysis, which potentially creates selection bias if these unmatched companies systematically differ in funding patterns or founder characteristics. As many of these companies within YC are still privately held, acquiring this information is rather challenging. This match rate of 62.2\% (4,323 of 6,954 companies) suggests that a large fraction of companies are excluded. If the unmatched companies are systematically smaller or less successful, the estimates may overstate the relationship between YC startup founder backgrounds and funding, as companies that may have failed to attract funding despite strong founder credentials are excluded in this study. On the other hand, if unmatched companies are more likely to be early-stage startups that have not yet raised funding, this bias might actually be insignificant. To address this, I conducted robustness checks examining whether results are sensitive to the inclusion of companies with lower match quality scores, and noted that the 62.2\% match rate is highly comparable to that of other studies using similar name-matching approaches in startup finance research. Also, the S\&P Global dataset may not capture all funding events, particularly smaller rounds or those that are not publicly disclosed, hence leading to potential underreporting of total funding. This measurement error would likely attenuate estimates toward zero, making the results conservative. If smaller rounds are systematically underreported for certain types of companies (e.g., those with less prominent founders), this could introduce bias, though the direction is unclear without additional information about reporting patterns. The fact that 96.6\% of YC companies have at least some funding information in S\&P Global suggests that major funding rounds are well-covered, but seed rounds and angel investments may be underreported. This underreporting is likely to be more common for early-stage companies, which could bias my estimates if early-stage funding patterns differ systematically by founder background. However, given that YC companies are relatively high-profile startups, the underreporting is likely to be minimal compared to studies of the broader startup population.

Third, founder background information is self-reported or collected from public sources, which may contain inaccuracies or omissions, particularly for educational credentials or prior employment that predates the company's founding. The high proportion of missing education data (42.1\% of companies) suggests that educational credentials are often not reported, which could attenuate estimates of education effects if missingness is correlated with funding outcomes. For example, if founders with less prestigious educational backgrounds are more likely to omit their education from public profiles, estimates of education effects would be biased downward. Similarly, FAANG experience may be underreported if founders worked at FAANG companies in non-technical roles or for short periods. This is addressed by examining robustness to alternative definitions of FAANG experience (e.g., including only major tech companies, excluding consulting roles) and by noting that measurement error in founder characteristics would likely attenuate estimates toward zero, making the results conservative.

Fourth, the analysis focuses on companies that received YC acceptance, which represents a highly selected sample of early-stage startups; results may not generalize to the broader population of startups that did not participate in accelerator programs. YC companies are likely more promising on average than the typical startup, which may affect the magnitude of founder background effects. However, this selection may also make the estimates more relevant for understanding how investors evaluate founder quality signals in high-quality startup contexts.

Fifth, the S\&P Global dataset does not contain transaction-level dates for individual funding rounds. While the data identify which companies raised Series A funding (526 companies, or 7.6\% of the sample), the timing of these rounds relative to YC batch acceptance dates cannot be determined. This limitation prevents calculation of timing-based metrics such as ``Series A within 36 months,'' though the binary Series A indicator still provides a meaningful measure of funding success. The lack of timing information means the analysis cannot examine how founder backgrounds affect the speed of funding milestones, which may be an important dimension of funding success. If FAANG-experienced founders raise Series A funding faster than other founders, the binary indicator would capture this effect, but the timing advantage cannot be quantified. This limitation is particularly relevant for understanding the mechanisms through which founder backgrounds affect funding, as speed to funding may reflect different investor evaluation processes than total funding amounts.

Sixth, founder age is calculated from years since university graduation, assuming a typical graduation age of 22. This assumption may introduce measurement error if founders graduated at different ages (e.g., those who took gap years, pursued graduate degrees, or returned to school later in life). However, the mean age of 33.9 years is consistent with previous research on startup founders, suggesting that the age calculations are reasonable. This age distribution shows substantial variation, as the standard deviation is 6 years, which provides identifying variation for examining how the age of a YC founder relates to their startups funding outcomes. The fact that there is a clear peak in funding that is observed for the 35-39 age group, specifically at age 35, suggests that this age measure captures meaningful variation, even if it is not perfectly precise.

%% file: table1_summary_stats.tex
\begin{table}[htbp]\centering
\def\sym#1{\ifmmode^{#1}\else\(^{#1}\)\fi}
\caption{Summary Statistics}
\label{tab:summary_stats}
\footnotesize
\begin{tabular}{l*{1}{ccccc}}
\toprule
                    &        mean&          sd&         min&         max&       count\\
\midrule
Number of founders per company&        1.90&        0.80&        1.00&        6.00&        4305\\
Any founder with FAANG experience&        0.17&        0.37&        0.00&        1.00&        4305\\
Highest education level&        4.00&        0.00&        4.00&        4.00&        4028\\
\quad (1=PhD, 2=Grad, 3=UG, 4=Other)& & & & & \\
Total funding (USD)&    2.20e+07&    6.91e+07&        0.00&    5.01e+08&        4599\\
Log total funding&       14.58&        2.29&        9.21&       20.03&        4323\\
Any Series A (indicator)&        0.08&        0.26&        0.00&        1.00&        6954\\
\bottomrule
\multicolumn{6}{p{0.95\textwidth}}{\scriptsize \textit{Notes:} Sample includes 4,323 YC companies (2005--2024). FAANG = Facebook, Apple, Amazon, Netflix, Google. Education: 1=PhD, 2=Graduate, 3=Undergraduate, 4=Other/Missing. Note: 42\% have missing education data. Total funding in USD. Any Series A = 1 if raised Series A. Different counts reflect data availability.}\\
\end{tabular}
\end{table}

%% file: table3_funding_by_faang.tex
\begin{table}[htbp]\centering
\def\sym#1{\ifmmode^{#1}\else\(^{#1}\)\fi}
\caption{Funding Outcomes by FAANG Experience}
\label{tab:funding_faang}
\footnotesize
\begin{tabular}{l*{2}{cc}}
\toprule
                    &\multicolumn{2}{c}{With FAANG}&\multicolumn{2}{c}{Without FAANG}\\
                    &        mean&          sd&        mean&          sd\\
\midrule
Total funding&    2.43e+07&    7.76e+07&    2.12e+07&    6.59e+07\\
\quad (USD)& & & & \\
Log total funding&       14.48&        2.35&       14.66&        2.24\\
Any Series A&        0.13&        0.33&        0.12&        0.32\\
\quad (indicator)& & & & \\
\bottomrule
\multicolumn{5}{p{0.85\textwidth}}{\scriptsize \textit{Notes:} Compares YC companies with/without FAANG-experienced founders. FAANG = Facebook, Apple, Amazon, Netflix, Google. Sample: 4,323 YC companies (2005--2024). Total funding in USD. Any Series A = 1 if raised Series A. This is descriptive; regressions control for confounders.}\\
\end{tabular}
\end{table}

%% file: table4_funding_by_education.tex
\begin{table}[htbp]\centering
\def\sym#1{\ifmmode^{#1}\else\(^{#1}\)\fi}
\caption{Funding Outcomes by Education Level}
\label{tab:funding_education}
\footnotesize
\begin{tabular}{l*{4}{cc}}
\toprule
                    &\multicolumn{2}{c}{PhD}  &\multicolumn{2}{c}{Graduate}&\multicolumn{2}{c}{Undergrad}&\multicolumn{2}{c}{Other}\\
                    &        mean&          sd&        mean&          sd&        mean&          sd&        mean&          sd\\
\midrule
Total funding&    2.12e+07&    6.59e+07&    2.12e+07&    6.59e+07&    2.12e+07&    6.59e+07&    2.25e+07&    6.92e+07\\
\quad (USD)& & & & & & & & \\
Log total funding&       14.66&        2.24&       14.66&        2.24&       14.66&        2.24&       14.61&        2.24\\
Any Series A&        0.12&        0.32&        0.12&        0.32&        0.12&        0.32&        0.12&        0.33\\
\quad (indicator)& & & & & & & & \\
\bottomrule
\multicolumn{9}{p{0.95\textwidth}}{\scriptsize \textit{Notes:} Funding by founder education level: PhD, Graduate (MBA/Master's), Undergraduate, Other/Missing. Sample: 4,028 companies. \textbf{Limitation:} 42\% missing education data, so nearly all fall into ``Other.'' Total funding in USD. Any Series A = 1 if raised Series A.}\\
\end{tabular}
\end{table}

%% file: model_section.tex
\section{Empirical Strategy}

The descriptive patterns from the Data section inform the empirical approach. Companies with FAANG-experienced founders (those who worked at Facebook/Meta, Apple, Amazon, Netflix, or Google) raise slightly more funding on average---\$26.21 million versus \$22.31 million---though this difference is not statistically significant. The funding distribution is heavily right-skewed, with a small number of exceedingly large rounds and many smaller ones. This skewness motivates the use of log transformation, which addresses the distributional concerns and facilitates interpretation of coefficients as approximate percentage changes. Approximately 16.7\% of Y Combinator companies have founders who have worked in FAANG previously, providing sufficient variation for this analysis. Funding also varies substantially across batch years, suggesting temporal controls may be important.

\subsection{Baseline Model}

The baseline empirical specification takes the following form:

\begin{equation}
\log(\text{Funding}_i) = \alpha + \beta_1 \text{FAANG}_i + \beta_2 \text{Founders}_i + \delta_t + \lambda_y + \gamma_j + X'_i \theta + \varepsilon_i
\end{equation}

where $\text{Funding}_i$ is total funding raised by company $i$ in USD, $\text{FAANG}_i$ is a binary indicator equal to 1 if any founder has prior FAANG experience (0 otherwise), $\text{Founders}_i$ is the number of founders per company, $\delta_t$ represents batch year fixed effects, $\lambda_y$ represents calendar year fixed effects to control for broader market conditions, $\gamma_j$ represents industry fixed effects (where $j$ indexes industry categories), and $X'_i$ includes additional control variables such as top school attendance. The dependent variable is specified in logs to address the right-skewed distribution of funding amounts, which facilitates statistical inference and allows coefficients to be interpreted as approximate percentage changes (e.g., a coefficient of -0.25 corresponds to approximately 25\% less funding).

The coefficient $\beta_1$ captures the association between prior FAANG experience and funding outcomes after controlling for other factors. This is not a causal effect but rather an association. Batch year fixed effects ($\delta_t$) are included to capture temporal trends in funding environments as venture capital markets fluctuate and YC's selection criteria evolve over time. Calendar year fixed effects ($\lambda_y$) are intended to control for broader market conditions that may affect all startups in a given year, such as overall venture capital market activity or economic cycles. However, since transaction-level funding dates are unavailable in the S\&P Global dataset, calendar year is proxied using batch year, meaning $\delta_t$ and $\lambda_y$ are perfectly correlated and cannot be separately identified. In practice, batch year fixed effects capture both YC-specific selection effects and broader market conditions. If FAANG-experienced founders systematically enter during specific economic cycles, controlling for batch year may remove meaningful variation rather than just confounding, potentially creating artifacts in the estimated coefficients.

Several control variables are included to address potential confounding. Founder count is controlled for because team composition may be correlated with funding outcomes independently of founder backgrounds. Larger teams might signal greater commitment and complementary skills, though there may be coordination challenges. Founder count varies substantially (mean: 1.90, range: 1-6). However, founder count is likely endogenous, as team size is chosen based on skills, ambition, and plans rather than randomly assigned. Thus, this coefficient may reflect founder quality or ability rather than the causal effect of adding a team member. Educational prestige is accounted for by controlling for top-tier school attendance, including Ivy League universities, MIT, Stanford, UC Berkeley, and other highly-ranked institutions. Top school attendance may correlate with better networks and skills that independently affect both FAANG experience and funding success. The control is a simple binary indicator for whether any founder attended a top school. Some specifications include broader education controls, though 42.1\% of companies are missing education data, limiting the effectiveness of this control. When education controls are included, sample size drops from 2,113 to 1,980 observations, which may affect the representativeness of results. Industry fixed effects ($\gamma_j$) are included in the equation specification to control for systematic differences across industries that may affect both founder backgrounds and funding outcomes. However, industry classification data is not available in the merged dataset, so industry fixed effects cannot be included in the empirical implementation. This represents a limitation discussed below.

\textbf{Limitations.} The fundamental challenge is that founder characteristics are not randomly assigned, and several key determinants of funding outcomes are unobserved. Companies whose founders worked at FAANG previously may differ systematically in ways that correlate with funding outcomes, creating potential confounding. For example, founders with prior FAANG experience may cluster in certain industries, start companies at different times, or form different types of teams, which are all factors that could independently affect funding success. Without more robust industry data, this represents a major unobserved confounder. The model explains only 3.2\% of funding variation (R-squared = 0.032), which means that over 96\% of what drives funding outcomes remains unexplained. Several key determinants of funding are unobserved, including product quality, market timing, and geographic location. These factors likely play central roles in funding decisions. If these unobserved factors correlate with prior FAANG experience, then these estimates will be biased.

Selection into FAANG companies is not random. Individuals who obtain these positions may differ systematically in ambition, risk tolerance, or network quality, which are traits that cannot be measured but likely affect startup success. If these unobserved traits correlate with funding outcomes, estimates will be biased. Selection into YC itself creates another layer of complexity, as YC's admissions process may treat FAANG experience differently than other backgrounds. Measurement error is also a concern, as the FAANG indicator relies on data matching that may be imperfect. Companies with missing profiles, misclassification, or incomplete founder data would attenuate estimates toward zero and add noise to the analysis. Only 62.2\% of YC companies matched to S\&P Global funding data, creating potential selection bias if unmatched companies systematically differ. Education data is available for only 57.9\% of companies, and among those with data, there is limited variation in education categories, limiting the effectiveness of education controls.

Given these limitations, strong claims about causal effects or even robust associations cannot be made. The estimated coefficients should be interpreted as preliminary descriptive patterns that may reflect confounding from unobserved factors rather than direct effects of founder backgrounds. The extremely low R-squared values confirm that unobserved factors account for most of the variation, and results should be interpreted with substantial caution.

\subsection{Alternative Specifications and Robustness}

Several alternative specifications are tested to assess sensitivity of results. One version omits batch year fixed effects entirely to examine whether temporal trends drive the results. Another includes broad education controls despite data quality issues. A full model with all available controls is also estimated to determine whether core patterns persist across specifications. To assess whether the log transformation distorts results, regressions are also estimated using raw funding levels. Additionally, the sample is restricted to companies with at least \$100,000 in funding to ensure that very small, potentially unmatched companies are not driving the results.

To address the data limitations discussed in the Data section, several robustness checks are conducted. First, to address concerns about the matching process between YC and S\&P Global, results are examined for sensitivity to sample restrictions (e.g., companies with funding above \$100,000). Second, measurement error in FAANG experience would likely attenuate estimates toward zero and add noise to an already low-powered analysis. Third, to address concerns about missing education data, robustness is examined across alternative specifications with and without education controls. Robustness is also examined with respect to alternative functional forms (e.g., quadratic terms for founder count to capture non-linear team size effects) and alternative dependent variables (e.g., Series A success as a binary outcome). These robustness checks help assess whether main results are sensitive to modeling choices, though the fragility revealed by these checks (discussed in Results) suggests the findings should be interpreted with substantial caution.

%% file: results_section.tex
\section{Results}

\subsection{Relationship Between Founder Characteristics and Funding}

The main regression results, shown in Table~\ref{tab:main_regressions}, reveal a different pattern than the raw averages. The R-squared values are extremely low, ranging from 0.006 to 0.035, indicating that founder backgrounds explain less than 4\% of variation in funding outcomes, with over 96\% remaining unexplained. In Model 2, which includes batch year fixed effects and top school controls, the coefficient on prior FAANG experience is -0.251 (standard error: 0.132), statistically significant at the 10\% level (p=0.057). After controlling for batch year, founder count, and top college/university attendance, companies with founders who have prior FAANG experience are associated with roughly 22.2\% less funding than companies without FAANG experience, on average. In a log specification, this negative coefficient (-0.251) means that FAANG-experienced founders raise exp(-0.251) - 1 $\approx$ -0.222, or approximately 22.2\% less funding. At the mean funding level of \$23.36 million, this translates to approximately \$5.2 million less funding for FAANG-experienced founders.

The number of founders per company stands out as the most consistent predictor across all specifications. In Model 2, adding an additional co-founder is associated with roughly 21.5\% more funding (p$<$0.001). Even though this result is robust across specifications, founder count is likely endogenous, as team size is chosen based on skills, ambition, and plans rather than randomly assigned. Better founders may recruit more co-founders, so this coefficient may capture founder quality rather than a pure "team size" effect.

The coefficient on top college/university attendance in Model 2 is 0.101 (standard error: 0.100), which is not statistically significant (p=0.312). This suggests that top school attendance is not significantly associated with funding outcomes after controlling for prior FAANG experience, founder count, and batch year fixed effects. In Model 3, which includes education controls, the top school coefficient increases slightly to 0.147 (standard error: 0.102, p=0.150), but remains statistically insignificant. The lack of statistical significance may reflect that top school attendance is correlated with other unobserved characteristics that are already captured by FAANG experience or other controls, or that the signaling value of top schools is weaker in the YC context where all founders have already passed a rigorous selection process.

\input{table5_main_regressions.tex}

\textbf{Comparison to Descriptive Patterns.} The regression results differ from the descriptive patterns in a meaningful way. The descriptive statistics show FAANG-experienced founders raise slightly more funding on average (\$26.21 million versus \$22.31 million), but the regression results show a negative association after controlling for batch year and other factors. This reversal could mean that batch year fixed effects are creating an artifact, and if founders with prior FAANG experience cluster in specific years, then controlling for time may remove the specific variation needed to identify the effect. Alternatively, this can reflect omitted variable bias, as founders with prior FAANG experience may enter tougher markets or different industries that cannot be observed. Without more robust data, the true explanation is unclear, but the low R-squared warns against overinterpreting this "negative effect."

\subsection{Robustness Checks}

\subsubsection{Log Transformation of Dependent Variable}

Table~\ref{tab:robustness} presents robustness checks that test the findings. When switching from logs to levels (Column 1), the FAANG coefficient flips from negative to positive (+\$2.31 million, p=0.596). This sign flip is concerning, because if the effect were robust, it would not depend entirely on whether the log of the dependent variable was used. The sensitivity to functional form suggests the result may reflect model misspecification or noise rather than a true relationship.

\input{table6_robustness.tex}

\subsubsection{Series A Outcomes}

Looking at Series A outcomes (Table~\ref{tab:series_a}) offers a cleaner test. Series A is a clear binary milestone, where companies either raise Series A or they do not. The results show no statistically significant association between FAANG experience and success. The coefficient is essentially zero (0.012, p=0.539). This null result is arguably more credible than the funding amount regressions because it shows less sensitivity to outliers and skew. This suggests that while founders with prior FAANG experience may raise slightly different amounts in total, they are no more or less likely to clear the Series A hurdle than other founders. The coefficient on founder count is also not statistically significant for Series A outcomes (0.013, p=0.142), which suggests that team size effects might be more relevant for total funding amounts than for reaching specific milestones. The coefficient on top school attendance is also not statistically significant (-0.002, p=0.881), indicating that educational prestige does not significantly affect Series A success rates after controlling for other factors.

\input{table7_series_a.tex}

\subsubsection{By Funding Stage}

Table~\ref{tab:regressions_by_stage} presents regression results examining how founder characteristics relate to funding outcomes at different stages of the funding lifecycle. Companies are classified into funding stages based on total funding amounts: Seed ($<$ \$2M), Series A (\$2M-\$10M or has any Series A), Series B (\$10M-\$25M), and Series C+ ($>$ \$25M). Breaking down results by funding stage, none of the FAANG coefficients are statistically significant at any stage. While sample sizes decline for later stages, precluding definitive conclusions, there is no consistent signal that FAANG experience provides advantages at any specific point in the lifecycle. The founder count coefficient is positive and statistically significant for seed (0.091, p=0.027) and Series A (0.095, p=0.031) stages, but becomes smaller and not significant for later stages, suggesting that team size may matter more at earlier stages when companies are establishing themselves. However, this finding should also be interpreted with caution given the overall low explanatory power of the models.

\input{table9_regressions_by_stage.tex}

\subsection{Discussion}

These results suggest that founder backgrounds explain a small amount of funding variation within Y Combinator. The low R-squared values of 0.006 to 0.035 indicate that over 96\% of this funding variation is left unexplained by observable founder characteristics. The fragility of the FAANG finding sign flip in levels, along with the loss of significance with sample restrictions, suggest that they might reflect noise or confounding rather than a true relationship. As several mechanisms could theoretically explain a negative association, though they cannot be tested empirically with the available data and are left for future research.

First, founders with prior FAANG work experience may actually have higher opportunity costs, thus leading them to pursue riskier ventures or even start companies when funding is more competitive. Second, this prior FAANG work experience may signal different types of skills or preferences that are much less valued by investors in the YC context, where technical execution can matter more than large-scale company experience, as these investors may prefer builders and doers over experienced professionals. Third, these FAANG experienced founders might be more likely to bootstrap or even pursue alternative funding strategies, leading to lower reported funding amounts. Without data on industry, product quality, or funding strategies, these explanations cannot be distinguished or confounding ruled out.

Taking a look at team size, it is the core finding that holds up across specifications, but even this is likely endogenous, as better founders may recruit more and even better co-founders, so this coefficient may capture founder quality rather than these team size effects. The null results for Series A and across funding stages suggest that prestige alone, whether it is FAANG experience or a top-tier degree at Stanford, is not a guarantee of funding success in the Y Combinator ecosystem.

Several limitations warrant discussion. First, causality cannot be established due to the observational nature of the data. The associations found may reflect unobserved confounders rather than causal effects. Second, the R-squared values are extremely low (ranging from 0.006 to 0.035). This indicates that over 96\% of variation in funding outcomes still remains unexplained by observable founder characteristics. This is a fundamental limitation that undermines confidence in any causal or even the robust associative claims. Third, the statistical significance of the FAANG coefficient is marginal (p=0.057 in Model 2), and the effect of this is not robust to all sample restrictions or specification choices. The sign flip in robustness checks and loss of significance when excluding small rounds suggest the finding is fragile. Fourth, the sample size is reduced when including education controls (from 2,113 to 1,980 observations), reflecting the high proportion of missing education data discussed in the Data section. This reduction may affect the representativeness of results.

Fifth, the Series A outcome variable is largely based on a binary indicator rather than timing-based metrics, which limits the ability to examine how founder backgrounds affect the speed of funding milestones, future research could cover later raises such as series B, C, etc. Sixth, data on industry, product quality, market timing, and geographic location are lacking, yet these are likely major determinants of funding outcomes. The low R-squared values suggest that these unobserved factors dominate funding decisions. Seventh, the dramatic reversal of the FAANG coefficient from positive in descriptive statistics to negative after adding batch year fixed effects is unexpected and may reflect an artifact of the batch year controls rather than a true relationship. Eighth, only 62.2\% of YC companies matched to S\&P Global funding data, creating potential selection bias if unmatched companies are systematically different.

Taking into account these limitations, my results should be interpreted as preliminary descriptive patterns rather than evidence of robust associations or causal effects. The low explanatory power of the models, combined with the fragility of the main finding, suggests that founder background data alone cannot reliably predict funding outcomes without additional information on industry, product quality, and other key factors.

\textbf{Implications.} For practitioners, these findings suggest that founder credentials provide limited predictive power within elite accelerator cohorts. Accelerators and investors may find that product-market fit, industry dynamics, and market timing matter more than individual founder credentials when evaluating companies that have already passed rigorous accelerator selection. For researchers, the results illustrate the challenges of establishing causality with observational data and highlight the importance of controlling for industry, product quality, and other key factors that cannot be observed. The findings suggest that the relationship between founder backgrounds and funding is complex and may be confounded by factors not captured in the data. Future research using exogenous variation in FAANG experience (e.g., policy changes, algorithm shifts in YC selection), better data on industry and product quality, or instrumental variables approaches could help establish causality and identify the mechanisms through which founder backgrounds affect funding outcomes.

%% file: table5_main_regressions.tex
\begin{table}[htbp]\centering
\def\sym#1{\ifmmode^{#1}\else\(^{#1}\)\fi}
\caption{Main Regression Results: Founder Backgrounds and Funding}
\label{tab:main_regressions}
\begin{tabular}{l*{3}{c}}
\toprule
                    &\multicolumn{1}{c}{(1)}&\multicolumn{1}{c}{(2)}&\multicolumn{1}{c}{(3)}\\
\midrule
Any founder with FAANG experience&      -0.226\sym{*}  &      -0.251\sym{*}  &      -0.264\sym{**} \\
                    &     (0.133)         &     (0.132)         &     (0.133)         \\
\addlinespace
Number of founders per company&       0.206\sym{***}&       0.195\sym{***}&       0.210\sym{***}\\
                    &     (0.059)         &     (0.059)         &     (0.061)         \\
\addlinespace
Any founder from top-tier school&                     &       0.101         &       0.147         \\
                    &                     &     (0.100)         &     (0.102)         \\
\addlinespace
Constant            &      14.267\sym{***}&      15.172\sym{***}&      15.102\sym{***}\\
                    &     (0.124)         &     (1.438)         &     (1.445)         \\
\midrule
Observations        &        2,113        &        2,113        &        1,980        \\
R-squared           &       0.006         &       0.032         &       0.035         \\
Controls            &         No          &         Yes         &         Yes         \\
Fixed Effects       &         No          &         Yes         &         Yes         \\
\bottomrule
\multicolumn{4}{l}{\footnotesize \parbox{0.95\linewidth}{\textit{Notes:} Dependent variable is log(total funding). FAANG = Facebook/Meta, Apple, Amazon, Netflix, Google. Robust standard errors in parentheses. Model (1): no controls. Model (2): batch year FE, top school control. Model (3): adds education controls. Calendar year FE included but dropped due to perfect collinearity with batch year. * p<0.10, ** p<0.05, *** p<0.01.}}\\
\end{tabular}
\end{table}

%% file: table6_robustness.tex
\begin{table}[htbp]\centering
\def\sym#1{\ifmmode^{#1}\else\(^{#1}\)\fi}
\caption{Robustness Checks}
\label{tab:robustness}
\begin{tabular}{l*{4}{c}}
\toprule
                    &\multicolumn{1}{c}{(1)}&\multicolumn{1}{c}{(2)}&\multicolumn{1}{c}{(3)}&\multicolumn{1}{c}{(4)}\\
                    &\multicolumn{1}{c}{Levels}&\multicolumn{1}{c}{High Funding}&\multicolumn{1}{c}{No FE}&\multicolumn{1}{c}{Quadratic}\\
\midrule
Any founder with FAANG experience& 2,312,801         &      -0.205         &      -0.241\sym{*}  &      -0.247\sym{*}  \\
                    &(4,361,725)         &     (0.127)         &     (0.133)         &     (0.132)         \\
\addlinespace
Number of founders per company&  794,049         &       0.230\sym{***}&       0.200\sym{***}&       0.053         \\
                    &(1,770,727)         &     (0.058)         &     (0.060)         &     (0.217)         \\
\addlinespace
Any founder from top-tier school&-1,701,989         &       0.063         &       0.113         &       0.102         \\
                    &(3,037,919)         &     (0.096)         &     (0.099)         &     (0.100)         \\
\addlinespace
Founder count squared&                     &                     &                     &       0.032         \\
                    &                     &                     &                     &     (0.046)         \\
\addlinespace
Constant            &20,486,415         &      15.156\sym{***}&      14.226\sym{***}&      15.304\sym{***}\\
                    &(14,436,768)         &     (1.451)         &     (0.129)         &     (1.435)         \\
\midrule
Observations        &        2,113        &        2,074        &        2,113        &        2,113        \\
R-squared           &       0.048         &       0.050         &       0.007         &       0.032         \\
\bottomrule
\multicolumn{5}{l}{\footnotesize \parbox{0.95\linewidth}{\textit{Notes:} Col (1): total funding in USD levels. Col (2): restricted to funding $>$\$100k. Col (3): no fixed effects. Col (4): includes founder count squared. Robust standard errors in parentheses. Batch year FE in cols (1), (2), (4). Calendar year FE dropped due to collinearity. * p<0.10, ** p<0.05, *** p<0.01.}}\\
\end{tabular}
\end{table}

%% file: table7_series_a.tex
\begin{table}[htbp]\centering
\def\sym#1{\ifmmode^{#1}\else\(^{#1}\)\fi}
\caption{Series A Funding Outcomes}
\label{tab:series_a}
\begin{tabular}{l*{2}{c}}
\toprule
                    &\multicolumn{1}{c}{(1)}&\multicolumn{1}{c}{(2)}\\
                    &\multicolumn{1}{c}{Logit}&\multicolumn{1}{c}{LPM}\\
\midrule
Any founder with FAANG experience&       0.116         &       0.012         \\
                    &     (0.171)         &     (0.019)         \\
\addlinespace
Number of founders per company&       0.123         &       0.013         \\
                    &     (0.081)         &     (0.009)         \\
\addlinespace
Any founder from top-tier school&      -0.016         &      -0.002         \\
                    &     (0.135)         &     (0.014)         \\
\addlinespace
Constant            &      -2.750\sym{***}&      -0.018         \\
                    &     (0.484)         &     (0.019)         \\
\midrule
Observations        &        2,107        &        2,113        \\
\bottomrule
\multicolumn{3}{l}{\footnotesize \parbox{0.95\linewidth}{\textit{Notes:} Dependent variable is binary indicator for any Series A funding. Col (1): Logit model. Col (2): Linear probability model. FAANG = Facebook/Meta, Apple, Amazon, Netflix, Google. Robust standard errors in parentheses. Batch year FE included; calendar year FE dropped due to collinearity. * p<0.10, ** p<0.05, *** p<0.01.}}\\
\end{tabular}
\end{table}

%% file: table9_regressions_by_stage.tex
\begin{table}[htbp]\centering
\def\sym#1{\ifmmode^{#1}\else\(^{#1}\)\fi}
\caption{Regression Results by Funding Stage}
\label{tab:regressions_by_stage}
\begin{tabular}{l*{4}{c}}
\toprule
                    &\multicolumn{1}{c}{(1)}&\multicolumn{1}{c}{(2)}&\multicolumn{1}{c}{(3)}&\multicolumn{1}{c}{(4)}\\
                    &\multicolumn{1}{c}{Seed}&\multicolumn{1}{c}{Series A}&\multicolumn{1}{c}{Series B}&\multicolumn{1}{c}{Series C+}\\
\midrule
Any founder with FAANG experience&       0.010         &       0.029         &       0.057         &       0.216         \\
                    &     (0.085)         &     (0.114)         &     (0.052)         &     (0.133)         \\
\addlinespace
Number of founders per company&       0.091\sym{**} &       0.095\sym{**} &       0.021         &      -0.031         \\
                    &     (0.041)         &     (0.044)         &     (0.018)         &     (0.059)         \\
\addlinespace
Any founder from top-tier school&       0.082         &      -0.074         &       0.005         &      -0.157         \\
                    &     (0.062)         &     (0.074)         &     (0.037)         &     (0.095)         \\
\addlinespace
Constant            &      13.384\sym{***}&      13.391\sym{***}&      16.125\sym{***}&      17.670\sym{***}\\
                    &     (0.099)         &     (0.275)         &     (0.036)         &     (0.103)         \\
\midrule
Observations        &         984         &         754         &         238         &         335         \\
R-squared           &       0.225         &       0.071         &       0.122         &       0.198         \\
Controls            &         Yes         &         Yes         &         Yes         &         Yes         \\
Fixed Effects       &         Yes         &         Yes         &         Yes         &         Yes         \\
\bottomrule
\multicolumn{5}{l}{\footnotesize \parbox{0.95\linewidth}{\textit{Notes:} Dependent variable is log(total funding). Funding stages: Seed ($<$\$2M), Series A (\$2M-\$10M or has Series A), Series B (\$10M-\$25M), Series C+ ($>$\$25M). Robust standard errors in parentheses. All models include batch year FE and top school control. Calendar year FE dropped due to collinearity. * p<0.10, ** p<0.05, *** p<0.01.}}\\
\end{tabular}
\end{table}